\documentclass[onefignum,onetabnum]{siamart171218}

\usepackage{lipsum}
\usepackage{amsfonts}
\usepackage{graphicx}
\usepackage{epstopdf}
\usepackage{algorithmic}
\usepackage{cite}
\usepackage{multirow}
\usepackage{hyperref}
\ifpdf
  \DeclareGraphicsExtensions{.eps,.pdf,.png,.jpg}
\else
  \DeclareGraphicsExtensions{.eps}
\fi

\newcommand\halfwidth{.46\textwidth}
\newcommand{\homdim}{p}

\newsiamremark{remark}{Remark}
\newsiamremark{hypothesis}{Hypothesis}
\crefname{hypothesis}{Hypothesis}{Hypotheses}
\newsiamthm{claim}{Claim}

\headers{Persistent Homology for Resource Coverage}{A. Hickok, B. Jarman, M. Johnson, J. Luo, M. A. Porter}

\title{Persistent Homology for Resource Coverage: A Case Study of Access to Polling Sites
\thanks{{AH, BJ, MJ, and JL contributed equally.}
\funding{AH was funded by the Eugene V. Cota-Robles fellowship, BJ was funded by NSF  grant number 2011140, and JL was funded by NSF grant number 1829071. {JL and MAP were funded by NSF grant number 1922952 through the Algorithms for Threat Detection (ATD) program.}
}}}

\author{Abigail Hickok\thanks{Department of Mathematics, Columbia University, NY, USA (\email{ah3966@columbia.edu}). AH worked on this project predominantly as a member of the Department of Mathematics at University of California, Los Angeles.}
\and Benjamin Jarman\thanks{Department of Mathematics, University of California, Los Angeles, CA, USA (\email{bjarman@math.ucla.edu}).} 
\and Michael Johnson\thanks{Department of Mathematics, University of California, Los Angeles, CA, USA (\email{mcjcard@math.ucla.edu}).} 
\and Jiajie Luo\thanks{Department of Mathematics, University of California, Los Angeles, CA, USA (\email{jerryluo8@math.ucla.edu}).} 
\and Mason A. Porter\thanks{Department of Mathematics, University of California, Los Angeles, CA, USA; Santa Fe Institute, Santa Fe, NM, USA (\email{mason@math.ucla.edu}).}
}

\usepackage{amsopn}

\makeatletter
\def\namedlabel#1#2{\begingroup
    #2%
    \def\@currentlabel{#2}%
    \phantomsection\label{#1}\endgroup
}
\makeatother

\makeatletter
\newcommand*{\addFileDependency}[1]{
  \typeout{(#1)}
  \@addtofilelist{#1}
  \IfFileExists{#1}{}{\typeout{No file #1.}}
}
\makeatother

\ifpdf
\hypersetup{
  pdftitle={Persistent Homology for Resource Coverage: A Case Study of Access to Polling Sites},
  pdfauthor={A. Hickok, B. Jarman, M. Johnson, J. Luo, and M. A. Porter}
}
\fi

\usepackage{subfig}
\usepackage{xurl}

\newcommand{\K}{\mathcal{K}}
\newcommand{\major}{} 

\begin{document}

\maketitle

\begin{abstract}
It is important to choose the geographical distributions of public resources in a fair and equitable manner. However, it is complicated to quantify the equity of such a distribution; important factors include distances to resource sites, availability of transportation, and ease of travel. We use persistent homology, which is a tool from topological data analysis, to study the effective availability and coverage of polling sites. The information from persistent homology allows us to infer holes in the distribution of polling sites. We analyze and compare the coverage of polling sites in Los Angeles County and five cities (Atlanta, Chicago, Jacksonville, New York City, and Salt Lake City), and we conclude that computation of persistent homology appears to be a reasonable approach to analyzing resource coverage.
\end{abstract}






\section{Introduction}

The geographical distribution of resources such as polling sites (i.e., locations where people vote), hospitals, COVID-19 vaccination sites, Department of Motor Vehicles (DMV) locations, and Planned Parenthood clinics is a major factor in the equitability of access to those resources. Consequently, given the locations of a set of resource sites, it is important to quantify their geographical coverage and to identify underserved geographical regions (i.e., ``holes in coverage'').

A naive approach to quantifying resource coverage is to consider the geographical distances from resource sites by simply calculating the percentage of people who reside within some cutoff distance $D$ of the nearest resource site. This naive approach is common in policy. For example, in March 2021, United States President Joseph Biden announced a goal to ensure that at least
90\% of the adult U.S. population is within $5$ miles (i.e., $D = 5$ miles) of a COVID-19 vaccination site \cite{whitehouse}. Additionally, it is required by Indian law that 100\% of voters live within $2$ km of a polling site \cite{india} (i.e., $D = 2$ km). However, such an approach poses at least two issues: 
\begin{enumerate}
    \item[\namedlabel{issue1}{(1)}] it requires choosing an arbitrary cutoff distance $D$; and
    \item[\namedlabel{issue2}{(2)}] using only geographical distance fails to account for many other factors, such as population density and the availability (and facility) of public transportation, that affect ease of access to a resource.
\end{enumerate}
These issues severely limit the utility of this naive approach.

In the present paper, we use \emph{topological data analysis} (TDA) to study holes in resource coverage. One of the main tools in TDA is \emph{persistent homology} (PH), which uses ideas from algebraic topology to (1) identify clusters and holes in a data set and (2) measure their persistences at different scales. We use PH to analyze data in the form of a \emph{point cloud}, which is a finite collection of points $X = \{x_i\}_{i=1}^n$ in a metric space $(M, d)$.\footnote{One can weaken the requirement that $d$ is a metric. In this paper, we use a distance function $d$ that is not technically a metric because it does not satisfy the triangle inequality.} In this paper, $X$ is a collection of resource sites, with specified latitudes and longitudes, and $M = \mathbb{R}^2$ with a non-Euclidean distance function
$d$ (see \cref{sec:methods}). Given a point cloud $X$ and a scale parameter $r > 0$, one can consider the \emph{$r$-coverage} $C_{r} := \bigcup_{i = 1}^n B(x_i, r)$. As the scale parameter $r$ grows, holes arise and subsequently fill in. PH tracks the formation and disappearance of these holes. When a point cloud is a collection of resource sites, one can interpret holes that persist for a large range of $r$ as holes in coverage. Our TDA approach gives a way to measure and evaluate how equitably a resource is distributed geographically.

Our approach using PH addresses both of the issues (see points \ref{issue1} and \ref{issue2}) of the naive approach that we discussed above.
First, PH eliminates the need to choose an arbitrary cutoff distance because one can study holes in coverage at all scales. 
Second, instead of employing geographical distance as our metric, we construct a distance function $d$ that is based on travel times. 
We also incorporate the waiting time at each resource site by constructing a weighted Vietoris--Rips (VR) filtration (see \cref{sec:background}) in which we weight vertices using estimates of waiting times at the associated sites. In a city with a high population density or a poor transportation system, the time that is spent waiting at or traveling to a resource site can be a much higher barrier to access than geographical distance \cite{votetraveltime, votetransportation}. We estimate waiting times by using Global Positioning System (GPS) ping data from mobile phones at the resource sites, and we estimate travel times by using street-network data, per capita car-ownership data, and the Google Maps application programming interface (API) \cite{googleapi}. Using these estimates, we construct a weighted VR filtration.
 We weight vertices by our estimates of waiting times, and we define the distance between two vertices to be the estimated round-trip travel time between them.
 Because the weighted VR filtration is stable, small errors in our estimates cause only small errors in the resultant PH \cite{weighted_VR}.

In the present paper, we examine polling sites as a case study of using PH to study the coverage of resource sites. We restrict our attention to six cities: Atlanta, Chicago, Jacksonville (in Florida), Los Angeles\footnote{For Los Angeles, we actually study Los Angeles County. We discuss the reasons for this choice in \cref{sec:la_nyc}.}, New York City (NYC), and Salt Lake City. We use these cities in part because data about them (e.g., car-ownership data) is widely available. Additionally, these cities differ considerably in their demographics and infrastructures, and we can thus compare a variety of different types of cities. Atlanta and NYC are both infamous for long waiting times at polling sites, especially in non-White neighborhoods \cite{atl_npr, nyc_intro}. In $2020$, some counties in the Atlanta metropolitan area had a mean of $3{\small,}600$ voters per polling site; the number of polling sites had been cut statewide by 10\% since $2013$ \cite{atl_npr}. In NYC, a mean of $4{\small,}173$ voters were assigned to each polling site in 2018. As a comparison, in $2004$, Los Angeles County and Chicago had only an estimated $1{\small,}300$ and $725$ voters per polling site, respectively \cite{nyc_intro}. However, Los Angeles is infamous for its traffic \cite{la_traffic}, which can affect voters' travel times to polling sites. Los Angeles and Chicago also differ in the quality of their public transportation, which also affects travel times to polling sites. In our investigation, we seek both to compare the coverage of polling sites in our six focal cities and to identify underserved areas within each city.


\subsection{Related work}

One can use tools from geography to study resource accessibility. Pearce et al.~\cite{Pearce2006-vf} used a geographical-information-systems (GIS) approach to examine the accessibility of community resources and how it affects health. Hawthorne and Kwan~\cite{Hawthorne2012-db} used a GIS approach and a notion of perceived distance to measure healthcare inequality in low-income urban communities. Brabyn and Barnett~\cite{Brabyn2004-hf} illustrated that there are regional variations in geographical accessibility to general practitioners in New Zealand and that these regional variations depend on how one measures accessibility.

Another motivation for our study of resource-site coverage is the related problem of sensor coverage. Given a set $S$ of sensors in a domain $\Omega \subseteq \mathbb{R}^2$, one seeks to determine if every point in $\Omega$ is within sensing range of at least one sensor in $S$. Typically, each sensor has a fixed, uniform sensing radius $r_s$. In this case, the problem is equivalent to determining whether or not the domain $\Omega$ is covered by balls of radius $r_s$ around each $s \in S$. In \cite{sensor, desilva2}, de Silva and Ghrist gave homological criteria for sensor coverage. Approaches to study sensor coverage that use computational geometry (specifically, ones that involve the Voronoi diagram of $S$ and the Delauney triangulation of $S$) were discussed in \cite{li, meguer}.

Our problem is also a coverage problem, but there are important differences. The key conceptual difference is that we consider neighborhoods whose sizes depend on a
filtration parameter, rather than neighborhoods of a fixed, uniform radius $r_s$. Additionally, we do not seek to determine whether or not the balls of any particular radius cover a
domain; instead, our goal is to quantify the coverage at all choices of radius and to determine how the holes in coverage evolve as we increase the filtration parameter. Another difference between the present paper and sensor-coverage problems is that our point cloud represents a set of resource sites (in particular, polling sites), rather than a set of sensors. In a sensor network, pairwise communication between sensors can play a role in whether or not the sensors are fully ``connected'' to each other (in a graph-theoretic sense) and in determining whether or not a domain is covered \cite{zhang2005}. By contrast, communication between resource sites does not play a role in access to those resource sites.

Several studies include applications of PH to geospatial data~\cite{corcoran_topological_2023}. Feng and Porter~\cite{feng2021} developed two methods to construct filtrations---one that uses adjacency structures and one that uses the level-set method \cite{osher2003} of front propagation---and applied their approaches to examine geospatial distributions of voting results in the 2016 United States presidential election. They identified ``political islands'' (i.e., precincts that voted more heavily for a candidate than the surrounding precincts). In \cite{feng2020}, Feng and Porter used their approaches to study spatial networks. Stolz et al.~\cite{brexit} used PH to examine the geospatial distribution of voting results in the ``Brexit'' referendum. Hickok et al.~\cite{covid} used PH to study geospatial anomalies in COVID-19 case-rate data (see also \cite{tda_spatial}) and vaccination-rate data. Cocoran and Jones \cite{corcoran_topological_2023} used PH to perform (1) a point-pattern analysis of pubs across different cities in the United Kingdom (UK) and (2) a spatiotemporal analysis of rainfall in the UK.


\subsection{Organization}

Our paper proceeds as follows. We present background information about PH in \cref{sec:background}, describe our approach in \cref{sec:methods}, present the results and analysis of persistence diagrams in \cref{sec:results}, and conclude and discuss implications, limitations, and potential future directions of our work in \cref{sec:discussion}. Our code is available at \url{https://bitbucket.org/jerryluo8/coveragetda/src/main/}.


\section{Background}\label{sec:background}

\begin{figure}
    \centering
    \subfloat[$\K_0$]{\includegraphics[width = .17\linewidth]{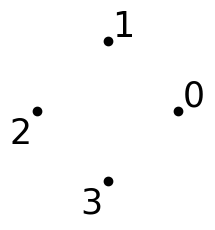}}
    \hspace{0.025\linewidth}
    \subfloat[$\K_1$]{\includegraphics[width = .17\linewidth]{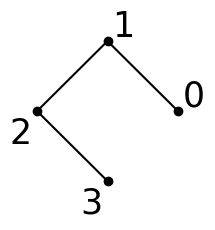}}
    \hspace{0.025\linewidth}
    \subfloat[$\K_2$]{\includegraphics[width = .17\linewidth]{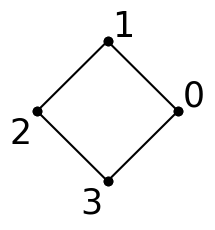}}
    \hspace{0.025\linewidth}
    \subfloat[$\K_3$]{\includegraphics[width = .17\linewidth]{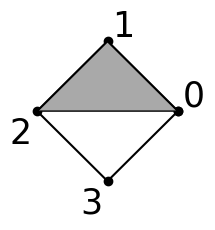}}
    \hspace{0.025\linewidth}
    \subfloat[$\K_4$]{\includegraphics[width = .17\linewidth]{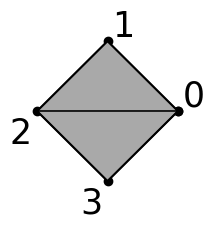}}
    \caption{An example of a filtration. The simplicial complex $\K_i$ has the associated filtration-parameter value $i$. [This figure appeared originally in \cite{covid}.]}
    \label{fig:fsc_example}
\end{figure}

We briefly review relevant mathematical background from 
\linebreak
TDA and PH. See \cite{roadmap,edel_book,dey_book} for more thorough discussions. To compute PH, we begin by constructing a ``filtered simplicial complex'' (which we will call a ``filtration'') from a point cloud $X$. A \emph{simplicial complex} is a combinatorial description of a topological space. It is a collection of vertices,
edges, triangles, and higher-dimensional \emph{simplices} with certain requirements on simplex boundaries and pairwise simplex intersections. A \emph{filtration} is a nested sequence $\K_{\alpha_0} \subseteq \K_{\alpha_1} \subseteq \cdots \subseteq \K_{\alpha_n}$ of simplicial complexes, where $\alpha_0 < \alpha_1 < \cdots < \alpha_n$. We show an example of a filtration in \cref{fig:fsc_example}.

\begin{figure}
    \centering
    \includegraphics[width = \linewidth]{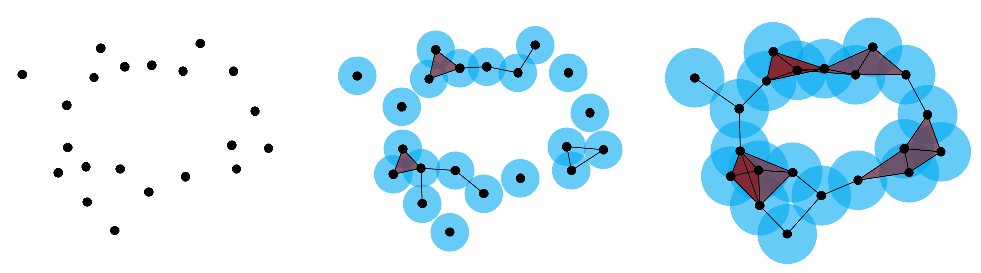}
    \caption{Illustration of a \v{C}ech filtration for a point cloud $X$ that we sample from an annulus. [We generated this figure using \cite{FiltrationDemos}.]}
    \label{fig:cech_example}
\end{figure}

Given a point cloud $X = \{x_1, \ldots, x_n\}$ in a metric space $(M, d)$, there are several ways to construct a filtration that approximates the shape of $X$. Two of the most common constructions are the \v{C}ech filtration and the Vietoris--Rips (VR) filtration \cite{roadmap}. For $r > 0$, the \emph{\v{C}ech complex $\mathrm{\check{C}}_{r}(X, M, d)$ at filtration parameter $r$} {is the simplicial complex that has a simplex with vertices $[x_{i_0}, \ldots, x_{i_k}]$ if the intersection $\bigcap_j B(x_{i_j}, r)$ is nonempty, where $B(x, r) := \{y \in M \mid d(x, y) \leq r\}$. (That is, $\mathrm{\check{C}}_{r}(X, M, d)$
is the nerve of the closed balls $\{B(x_i, r)\}_{x_i \in X}$.) By the Nerve Theorem \cite{Borsuk}, $\mathrm{\check{C}}_{r}(X, M, d)$ is topologically equivalent (more precisely, it is homotopy-equivalent)
to the union $\bigcup_i B(x_i, r)$ of balls (i.e., the ``$r$-coverage'' of $X$) in $M$ whenever the balls $B(x_i, r)$ are convex.\footnote{This condition is satisfied for all $r$ when $(M, d)$ is Euclidean, but it is not always satisfied for non-Euclidean metric spaces.} 
This implies that $\bigcup_i B(x_i, r)$ and $\mathrm{\check{C}}_{r}(X, M, d)$ have the same homology (i.e., the same set of holes). 
A \emph{\v{C}ech filtration} is a nested sequence of \v{C}ech complexes for increasing filtration parameter $r$. In \cref{fig:cech_example}, we show an example of a \v{C}ech filtration. 

In practice, it is uncommon to use \v{C}ech filtrations because they are} difficult to compute. A \emph{Vietoris--Rips (VR) complex} $\mathrm{VR}_{r}(X, M, d)$ is an approximation of a \v{C}ech complex that is faster to compute because it is only necessary to calculate pairwise distances between points. The VR complex at filtration parameter $r$ has a simplex with vertices $[x_{i_0}, \ldots, x_{i_k}]$ if $d(x_{i_j}, x_{i_{\ell}}) < 2 r$ for all $j$ and $\ell$. A \emph{VR filtration} is a nested sequence of VR complexes for a sequence of increasing filtration parameters. A VR filtration ``approximates'' a \v{C}ech filtration in the sense that
\begin{equation}\label{eq:vrlemma}
    \mathrm{\check{C}}_{r}(X, M, d) \subseteq \mathrm{VR}_{r}(X, m, d) \subseteq \mathrm{\check{C}}_{\sqrt{2}r}(X, M, d)
\end{equation}
for all $r$. The complexes $\mathrm{VR}_{r}(X, M, d)$ and $\mathrm{\check{C}}_r(X, M, d)$ have the same set of edges for all $r$.

Weighted versions of the \v{C}ech and VR filtrations were described in \cite{weighted_VR}. Given a point cloud $X = \{x_1, \ldots,x_n\}$ in a metric space $(M, d)$ and associated weights $\{w_1,\ldots,w_n\}$, the \emph{radius function} at $x_i$ is
\begin{equation}\label{eq:radius_function}
    r_{x_i}(t) := \begin{cases}
        -\infty \,, & t < w_i \\
        t - w_i\,, & \text{otherwise\,.}
    \end{cases}
\end{equation}
The closed ball $B(x_i, r_{x_i}(t))$ has no points until time $t = w_i$; at that time, the radius starts growing linearly with $t$, which is the filtration parameter. The \emph{weighted \v{C}ech complex} $\mathrm{\check{C}}^{\text{weighted}}_t(X, M, d, \{w_i\})$ at filtration parameter $t$
is the simplicial complex that has a simplex with vertices $[x_{i_0}, \ldots, x_{i_k}]$ if the intersection $\bigcap_j B(x_{i_j}, r_{x_{i_j}}(t))$ is nonempty. (That is, $\mathrm{\check{C}}^{\text{weighted}}_t(X, M, d, \{w_i\})$ is the nerve of $\{B(x_i, r_{x_i}(t))\}_{x_i \in X}$.) Like the unweighted \v{C}ech complex, the weighted \v{C}ech complex is homotopy-equivalent to the union $\bigcup_i B(x_i, r_{x_i}(t))$ of balls by the Nerve Theorem whenever the balls $B(x_i, r_{x_i}(t))$ are convex for all $x_i$. Much like an unweighted \v{C}ech complex, a weighted \v{C}ech complex requires too much time to compute in practice, so researchers usually instead compute a \emph{weighted VR complex} $\mathrm{VR}^{\text{weighted}}_t(X, M, d, \{w_i\})$. This is the simplicial complex whose vertices are $\{x_i \mid w_i < t\}$ and whose simplices $[x_{i_0}, \ldots, x_{i_k}]$ satisfy $d(x_{i_j}, x_{i_{\ell}}) + w_{i_j} + w_{i_\ell} < 2t$. The sequence $\{\mathrm{VR}^{\text{weighted}}_t(X, M, d, \{w_i\})\}_t$ for increasing $t$ is the \emph{weighted VR filtration}. Analogously to \cref{eq:vrlemma}, the weighted VR filtration ``approximates'' the weighted \v{C}ech filtration in the sense that
\begin{equation}\label{eq:weighted_vrlemma}
    \mathrm{\check{C}}_{r}(X, M, d, \{w_i\}) \subseteq \mathrm{VR}_{r}(X, m, d, \{w_i\}) \subseteq \mathrm{\check{C}}_{\sqrt{2}r}(X, M, d, \{w_i\})
\end{equation}
for all $r$.

Given a filtration $\K_{\alpha_0} \subseteq \cdots \subseteq \K_{\alpha_n}$, one can compute the \emph{homology} of each simplicial complex $\K_{\alpha_i}$. A \emph{homology class} represents a hole that exists in a filtration for some range of filtration-parameter values $\alpha_i$. A $0$-dimensional (0D) homology class represents a connected component and a $1$-dimensional (1D) homology class represents a hole that is bounded by a closed path. To see why 0D homology classes are ``holes'', we note that one can think of a 0D homology class as a representative of the empty space between connected components. Therefore, in identifying ``holes", it is important to consider both 0D and 1D homology classes.

As the filtration parameter $\alpha_i$ grows, holes form and subsequently fill in. The information that is given by the birth and death of {the homology classes} of a filtration is called the \emph{persistent homology} (PH) of the filtration. We say that a homology class is \emph{born} at $\alpha_i$ if $i$ is the minimum index such that the homology class appears in $K_{\alpha_i}$. Its \emph{birth simplex} is the simplex that creates the homology class. For example, in \cref{fig:fsc_example}, a 1D homology class is born at step $2$. Its birth simplex is the edge with vertices $0$ and $3$.
A homology class that is born at $\alpha_i$ subsequently \emph{dies} at $\alpha_j$, with $j \geq i$, if $j$ is the minimum index such that the homology class becomes trivial (i.e., the corresponding hole fills in)
in $K_{\alpha_j}$. We refer to $\alpha_i$ as the homology class's \emph{birth value} and to $\alpha_j$ as its \emph{death value}. Its \emph{death simplex} is the simplex that destroys the homology class. For example, the homology class that is born at step $2$ in \cref{fig:fsc_example} subsequently dies at step $4$. Its death simplex is the triangle with vertices $0$, $2$, and $3$.
In our application to polling sites, we interpret homology classes as holes in coverage and we interpret the death simplices as the locations of the holes.

\begin{figure}
    \centering
    \includegraphics[width = .5\linewidth]{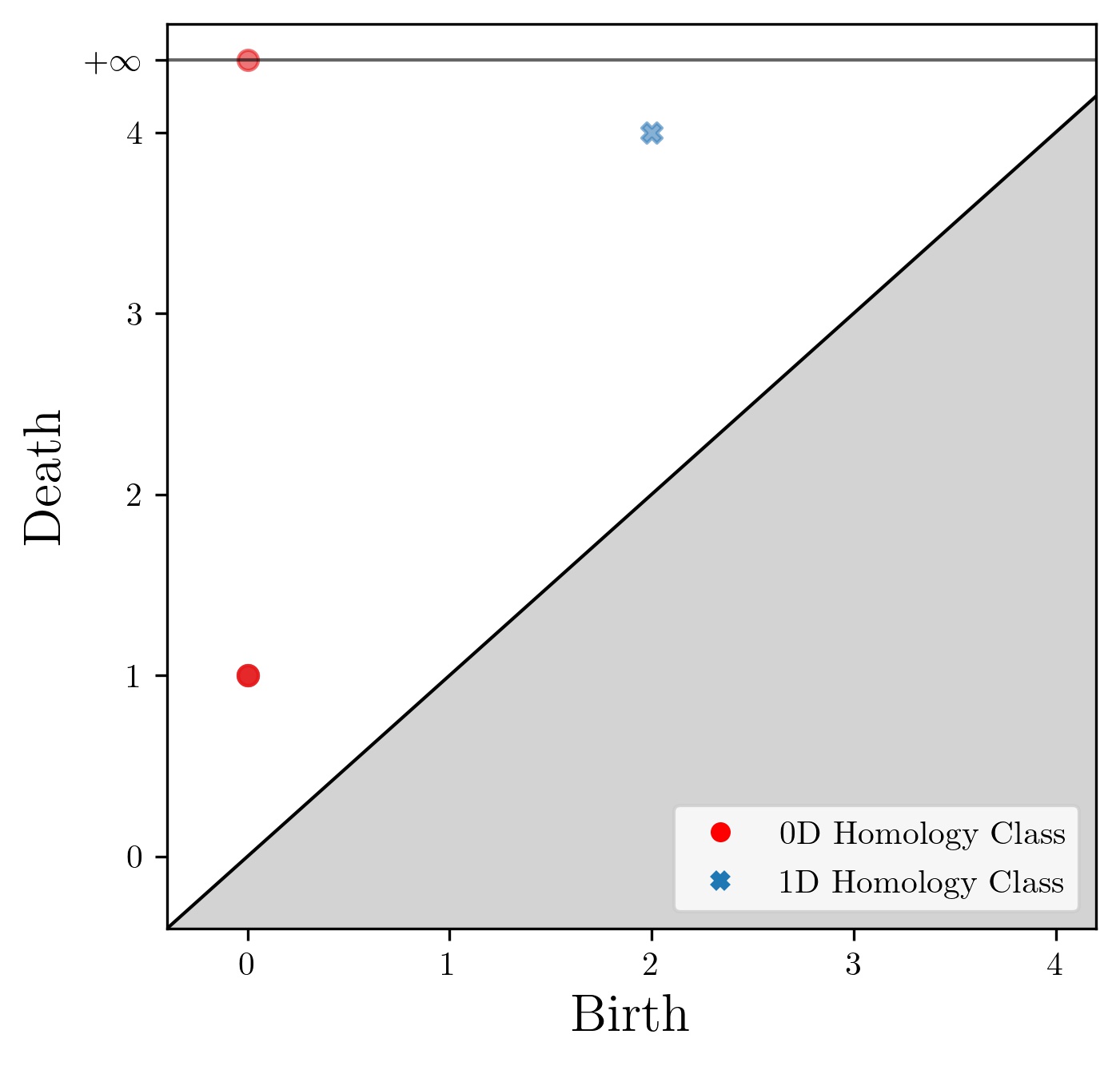}
    \caption{The persistence diagram for the 0D and 1D PH of the filtration in \cref{fig:fsc_example}.}
    \label{fig:pd_example}
\end{figure}

One can summarize PH in a \emph{persistence diagram} (PD). A PD is a multiset of points in the extended plane $\overline{\mathbb{R}}^2$. For a homology class with birth value $b$ and death value $d$, the PD includes a point with coordinates $(b, d)$. 
In \cref{fig:pd_example}, we show the PD for the PH of the filtration in \cref{fig:fsc_example}.

\section{Our Construction of Weighted VR Complexes}\label{sec:methods}

For each city, we construct a weighted VR filtration in which the point cloud $X = \{x_i\}$ is the set of polling sites in $\mathbb{R}^2$ and the weight $w_i$ of a point $x_i$ is an estimate of the waiting time at the corresponding polling site. Instead of computing a weighted VR filtration with respect to Euclidean distance, we define a distance function that estimates
the mean amount of time that it takes to travel to and from a polling site. With respect to this distance function, the union $\bigcup_i B(x_i, r_{x_i}(t))$ 
(see \cref{eq:radius_function}) is the set of points $y$ such that the estimated time for an individual at $y$ to vote (including waiting time and travel time\footnote{Incorporating information (such as waiting times) other than travel times is sensible both in principle and in practice. In our computational experiments, using only travel times yields results that differ drastically from those that we present in this paper.}
in both directions) is at most $t$.
The weighted \v{C}ech complex $\mathrm{\check{C}}_t^{\text{weighted}}(X, \mathbb{R}^2, d, \{w_i\})$ is an approximation of $\bigcup_i B(x_i, r_{x_i}(t))$. When the balls $B(x_i, r_{x_i}(t))$ are convex, the weighted \v{C}ech complex is homotopy-equivalent to $\bigcup_i B(x_i, r_{x_i}(t))$, so these two complexes have the same homology (i.e., the same set of holes). The weighted VR complex $\mathrm{VR}_t^{\text{weighted}}(X, \mathbb{R}^2, d, \{w_i\})$ is an approximation of the weighted \v{C}ech complex.

We construct our distance function as follows. Let $x$ and $y$ be two polling sites. We estimate the expected time for an individual to travel from $x$ to $y$ and back to be 
\begin{align*}
	\tilde{d}(x,y) &:= C(Z(x))\min\{t_{\mathrm{car}}(x,y), t_{\mathrm{pub}}(x,y), t_{\mathrm{walk}}(x,y)\}\\
&\qquad + [1 - C(Z(x))]\min\{t_{\mathrm{pub}}(x,y)\,, t_{\mathrm{walk}}(x,y)\}\,,
\end{align*}
where $Z(x)$ is the zip code that includes $x$ (a polling site), $C(Z(x))$ is our estimate of the fraction of voting-age people in $Z(x)$ who can travel by car to a polling site, and $t_{\mathrm{car}}(x,y)$, $t_{\mathrm{pub}}(x,y)$, and $t_{\mathrm{walk}}(x,y)$ are our estimates of the expected travel time from $x$ to $y$ and back by car, public transportation,
and walking, respectively. We estimate $C(Z(x))$ by dividing an estimate of the number of personal vehicles in $Z(x)$ by an estimate of the voting-age population in $Z(x)$; see \cref{sec:otherestimates}. We discuss how we calculate $t_{\mathrm{car}}$, $t_{\mathrm{pub}}$, and $t_{\mathrm{walk}}$ in \cref{sec:traveltime}.

Our definition of $\tilde{d}(x,y)$ captures the cost (in time) to vote. In particular, $\tilde{d}(x,y)$ is an estimate of the mean travel time for an individual
who resides in zip code $Z(x)$ to travel from $x$ to $y$ and back. We assume that all individuals choose the fastest mode of transportation that is available to them. Specifically, we assume that individuals who can travel by car choose the fastest option between driving, taking public transportation, and walking. That is, their travel time is $\min\{t_{\mathrm{car}}(x, y), t_{\mathrm{pub}}(x, y), t_{\mathrm{walk}}(x, y)\}$. Likewise, we assume that individuals who do not have access to travel by car choose the fastest option between taking public transportation and walking. That is, their travel time is $\min\{t_{\mathrm{pub}}(x, y), t_{\mathrm{walk}}(x,y)\}$. We estimate that the fraction of a population that has a car to be
$C(Z(x))$, so the fraction without a car is $1 - C(Z(x))$. Therefore, $\tilde{d}(x, y)$ is the (estimated) mean time for an individual
who resides in zip code $Z(x)$ to travel from $x$ to $y$ and back.

{\major{The function $\tilde{d}(x, y)$ is not symmetric (i.e., $\tilde{d}(x, y) \neq \tilde{d}(y, x)$) because $C(Z(x)) \neq C(Z(y))$.}} However, we need a symmetric function to construct a weighted VR filtration. To construct a 
symmetric distance function that is based on $\tilde{d}(x,y)$, we define the distance between $x$ and $y$ to be a weighted average of $\tilde{d}(x,y)$ and $\tilde{d}(y,x)$, where we determine the weights from the populations of the zip codes that include $x$ and $y$. More precisely, we define the distance between $x$ and $y$ to be
\begin{equation}\label{eq:distance}
    d(x, y) := \frac{1}{P}[P_{Z(x)}\tilde{d}(x, y) + P_{Z(y)} \tilde{d}(y, x)]\,,
\end{equation}
where $P_{Z(x)}$, and $P_{Z(y)}$ are the populations of
zip codes $Z(x)$ and $Z(y)$, respectively, and $P :=  P_{Z(x)}+ P_{Z(y)}$ is the 
sum of the populations of $Z(x)$ and $Z(y)$. With respect to this distance function, the ball $B(x, r)$ is the set of points $y$ such that the expected time for an individual to travel back and forth between $x$ and $y$ is at most $r$, where we randomly choose the individual
to start at either $x$ or $y$ with probabilities that are weighted by the populations of their associated zip codes. 
Although our distance function is not technically a metric (because it does not satisfy the triangle inequality), we can still construct a weighted VR filtration using the definition in
\cref{sec:background}.


\subsection{Estimating travel times}\label{sec:traveltime}

To compute our distance function (see \cref{eq:distance}), we need to estimate the pairwise travel times
by car, public transportation, 
and walking between each pair of polling sites. We measure these times in minutes.

\begin{figure}[h]
    \centering
    \includegraphics[width=\linewidth]{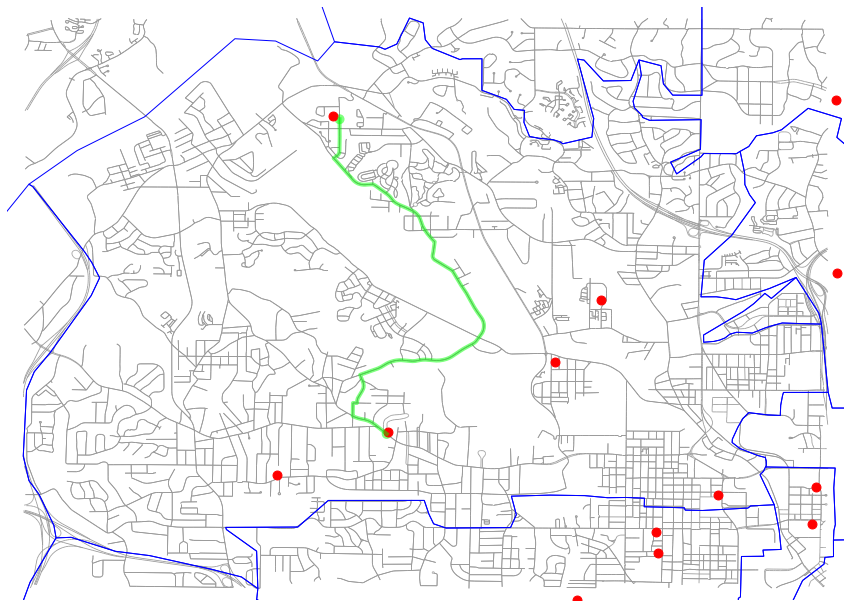}
    \caption{A shortest path (by geographical distance) between two polling sites in zip code 30314 in Atlanta.
    }
    \label{fig:path example}
\end{figure}

We estimate the time that it takes to walk between each pair of polling sites by using street networks, which are available through the OpenStreetMap tool \cite{openstreetmap}, for each of our
cities. Using OpenStreetMap, we calculate a shortest path (by geographical distance) between each pair of polling sites. In \Cref{fig:path example}, we show an example of a shortest path between two polling sites in Atlanta.

Let $L(x, y)$ denote the length (which we measure in meters) of a shortest path (by geographical distance) between polling sites $x$ and $y$. Our estimate of the walking time (in minutes) from $x$ to $y$ and back is $t_{\mathrm{walk}}(x, y) := 2L(x, y)/v_{\mathrm{walk}}$, where $v_{\mathrm{walk}} = 85.2$ meters per minute is an estimate of the mean 
walking speed of an adult human \cite{walk_speed}.

To estimate travel times by car and by public transportation, we use the Google Maps Distance Matrix API \cite{googleapi}. Because of budgetary constraints (and the cost of five dollars per thousand API queries), we 
use this API to estimate only the travel times between each polling site and its $25$ geographically closest polling sites. We refer to these sites as a polling site's $25$ nearest neighbors.

For each of the $25$ nearest neighbors, we separately calculate both the time \emph{from} a polling site \emph{to} each neighbor and the time \emph{to} a polling site \emph{from} each neighbor. These two travel times are often different because of different traffic conditions or other factors. We estimate the remaining pairwise travel times as follows. Let $G$ be the unweighted, undirected graph whose vertices are the polling sites and whose edges connect each vertex to its $25$ nearest geographical neighbors.\footnote{The relation of being one of a vertex's 25 nearest neighbors is not symmetric. Therefore, the degrees of some vertices are larger than $25$.} 
Let $G_{\mathrm{car}}$ and $G_{\mathrm{pub}}$ be the weighted, directed graphs whose vertices and edges\footnote{\major{We view each undirected edge $(x_i, x_j)$ of $G$ as a bidirectional edge, and we include both of the associated directed edges in the directed graphs $G_{\mathrm{car}}$ and $G_{\mathrm{pub}}$.}} are those of $G$ and whose weights are given by the travel times (by car and public transportation, respectively) that we compute using the Google Maps API. The weight of the directed edge from vertex 
$x$ to vertex $y$ is the travel time from $x$ to $y$. Therefore, the weight of the edge from $x$ to $y$ may differ from the weight of the edge from $y$ to $x$.
For any two polling sites $x$ and $y$, let the travel times $\tilde{t}_{\mathrm{car}}(x, y)$ and $\tilde{t}_{\mathrm{pub}}(x, y)$ be 
the length of a shortest weighted path from $x$ to $y$ in the graphs $G_{\mathrm{car}}$ and $G_{\mathrm{pub}}$, respectively. The corresponding symmetrized travel times $t_{\mathrm{car}}(x, y)$ and $t_{\mathrm{pub}}(x, y)$ are
\begin{align*}
    t_{\mathrm{car}}(x, y) &:= \tilde{t}_{\mathrm{car}}(x, y) + \tilde{t}_{\mathrm{car}}(y, x)\,, \\
    t_{\mathrm{pub}}(x, y) &:= \tilde{t}_{\mathrm{pub}}(x, y) + \tilde{t}_{\mathrm{pub}}(y, x)\,.
\end{align*}


\subsection{Estimating waiting times}\label{sec:waittime}

Our weighted VR filtrations have weights at each vertex (i.e., polling site) that are given by an estimate of the mean time that a voter spends (i.e., the mean waiting time)
at that polling site. In a nationwide study of waiting times at polling sites~\cite{waittimes}, Chen et al. used smartphone data of hundreds of thousands of voters to estimate waiting times. They also examined potential relationships between waiting times and racial demographics.

We construct our estimates using the congressional district-level estimates in \cite{waittimes} (see their Table C.2). For each polling site $x$, we take the mean of the waiting-time estimates for each congressional district that overlaps with the zip code $Z(x)$ that contains $x$. {This averaging procedure yields estimates of waiting times at
the zip-code level. (We transform our waiting-time data to the zip-code level because the rest of our data is at the zip-code level.)}


\subsection{Estimates of demographic information}\label{sec:otherestimates}

We obtain estimates of demographic data at the zip-code level from 2019 five-year American Community Survey data \cite{uscensus}. 
We use voting-age population data from their Table ACSDT5Y2019.B29001 and vehicle-access data from their Table~ACSDT5Y2019.B25046.


\subsection{Polling-site zip codes}\label{sec:preprocess}

Much of our data is at the zip-code level, and we treat a polling site's zip code as representative of its local area. Certain polling sites (predominantly government buildings) have their own zip codes, despite their populations of $0$. We adjust the zip codes of such polling sites to match the zip codes of the directly surrounding areas.


\subsection{Treatments of Los Angeles and New York City}\label{sec:la_nyc}

Because of the oddly-shaped city boundaries of Los Angeles, which include several holes, we use the entirety of Los Angeles County (except for its islands).

Because of the disconnected nature of New York City, we subdivide it into three regions (Queens and Brooklyn, Manhattan and the Bronx, and Staten Island) and treat each region separately. We then combine our results for the three regions into a single presentation. For example, we combine the PDs into a single PD for all of New York City.


\section{Results}\label{sec:results}

\begin{figure}
    \centering
    \subfloat[Atlanta]{\includegraphics[width=\halfwidth]{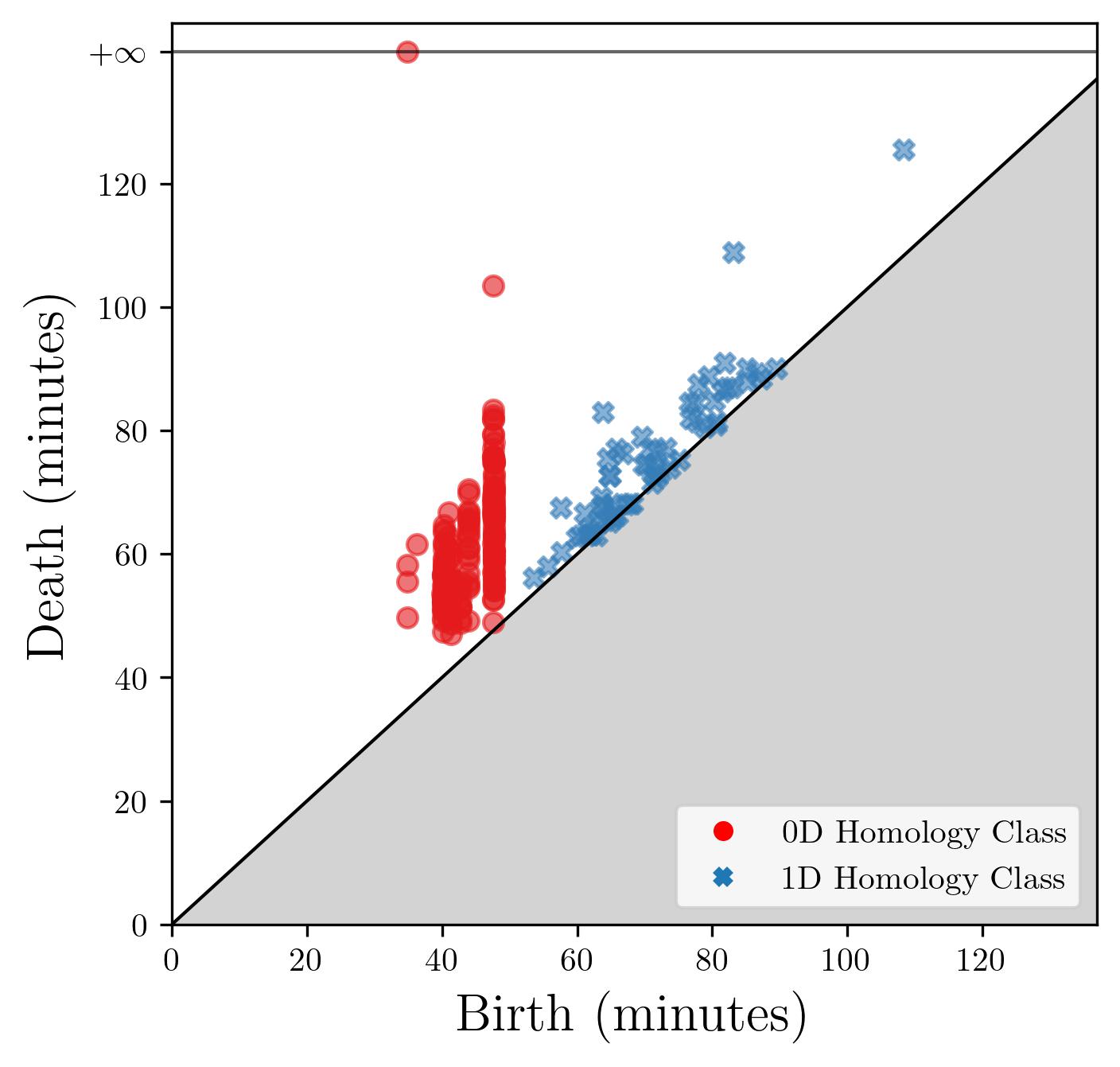}\label{fig:AtlantaPD}}
    \hspace{.07\textwidth}
    \subfloat[Chicago]{\includegraphics[width=\halfwidth]{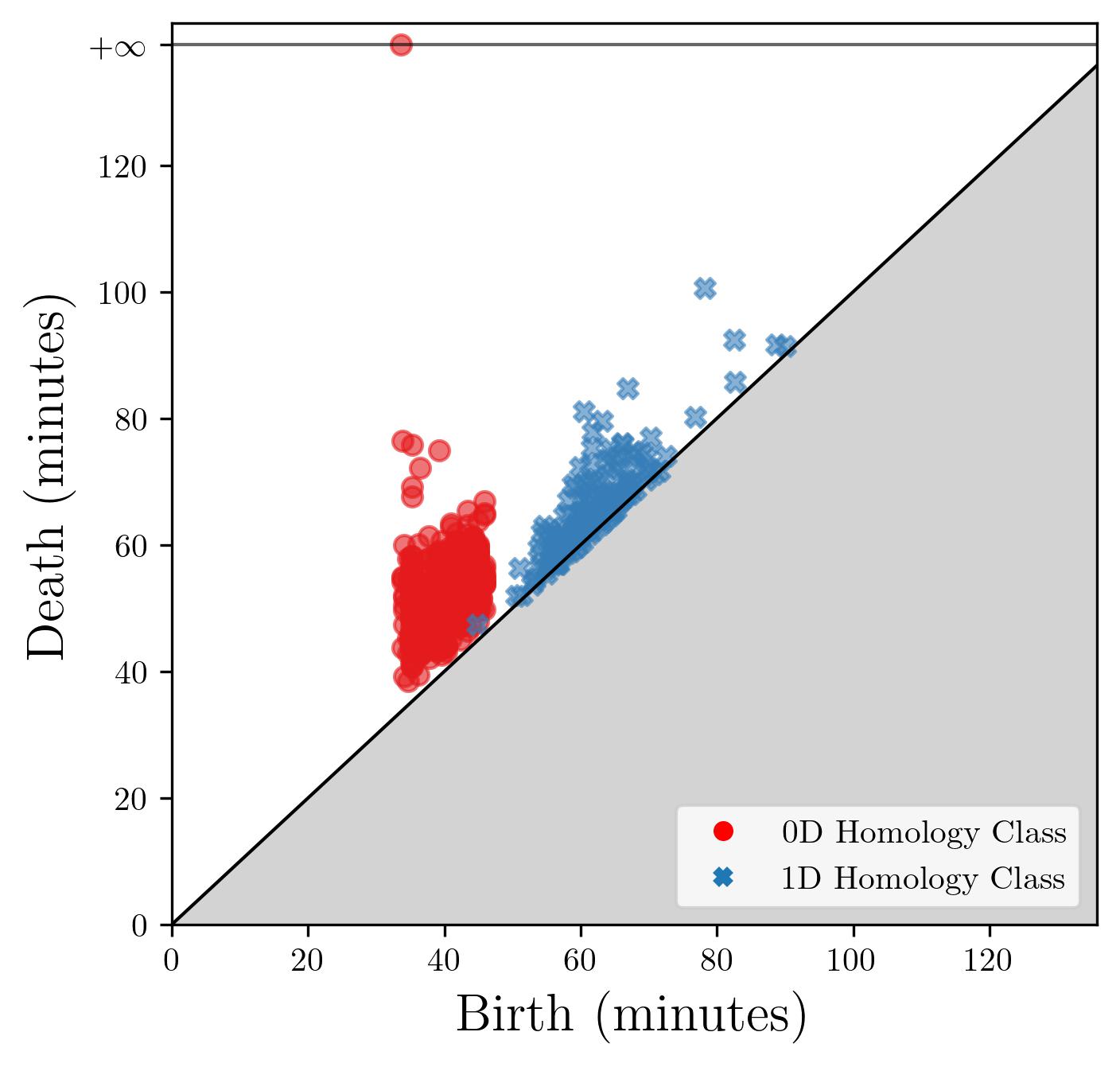}\label{fig:ChicagoPD}} \\
    \subfloat[Jacksonville (Florida)]{\includegraphics[width=\halfwidth]{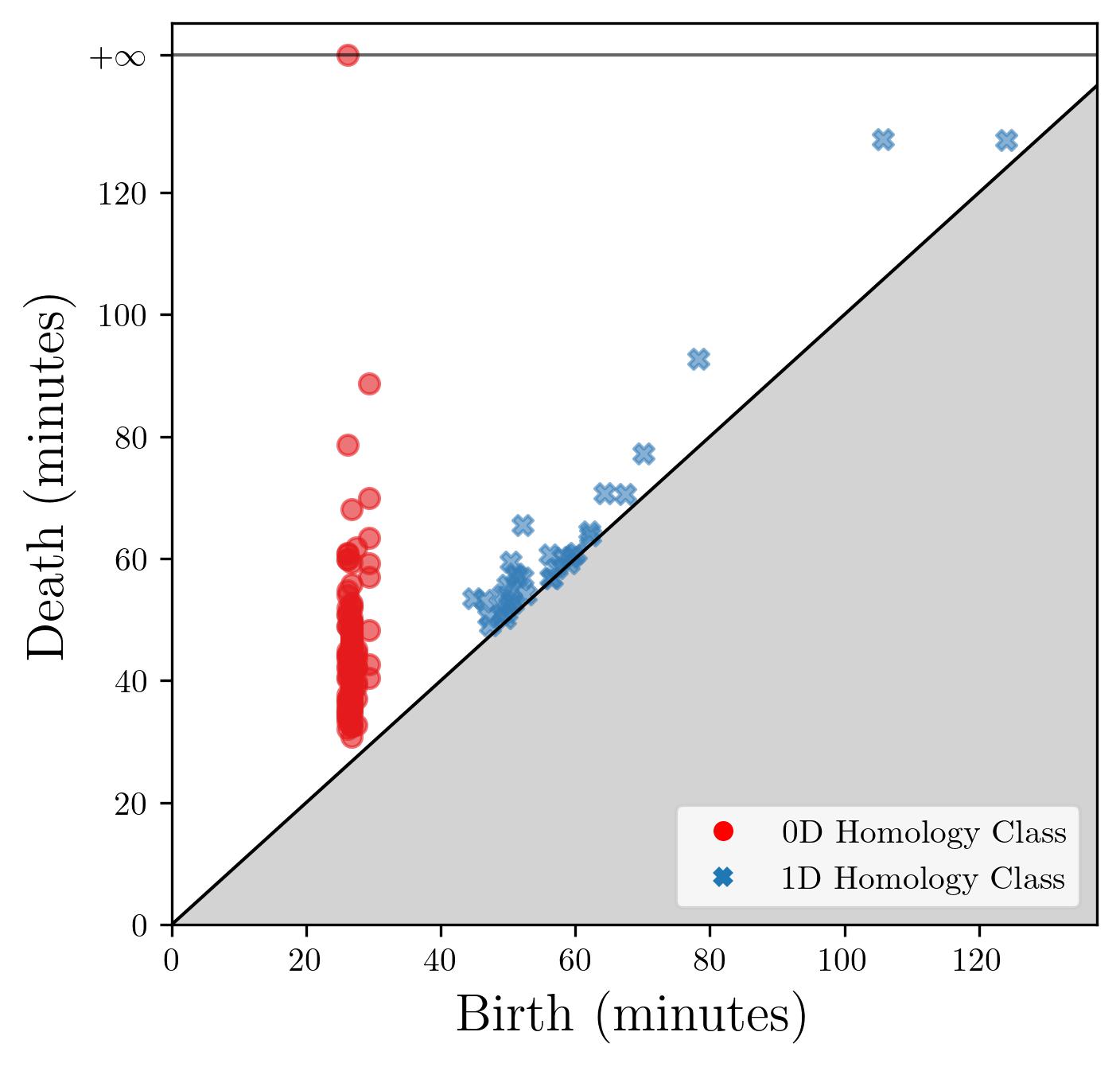}}
    \hspace{.07\textwidth}
    \subfloat[Los Angeles County]{\includegraphics[width=\halfwidth]{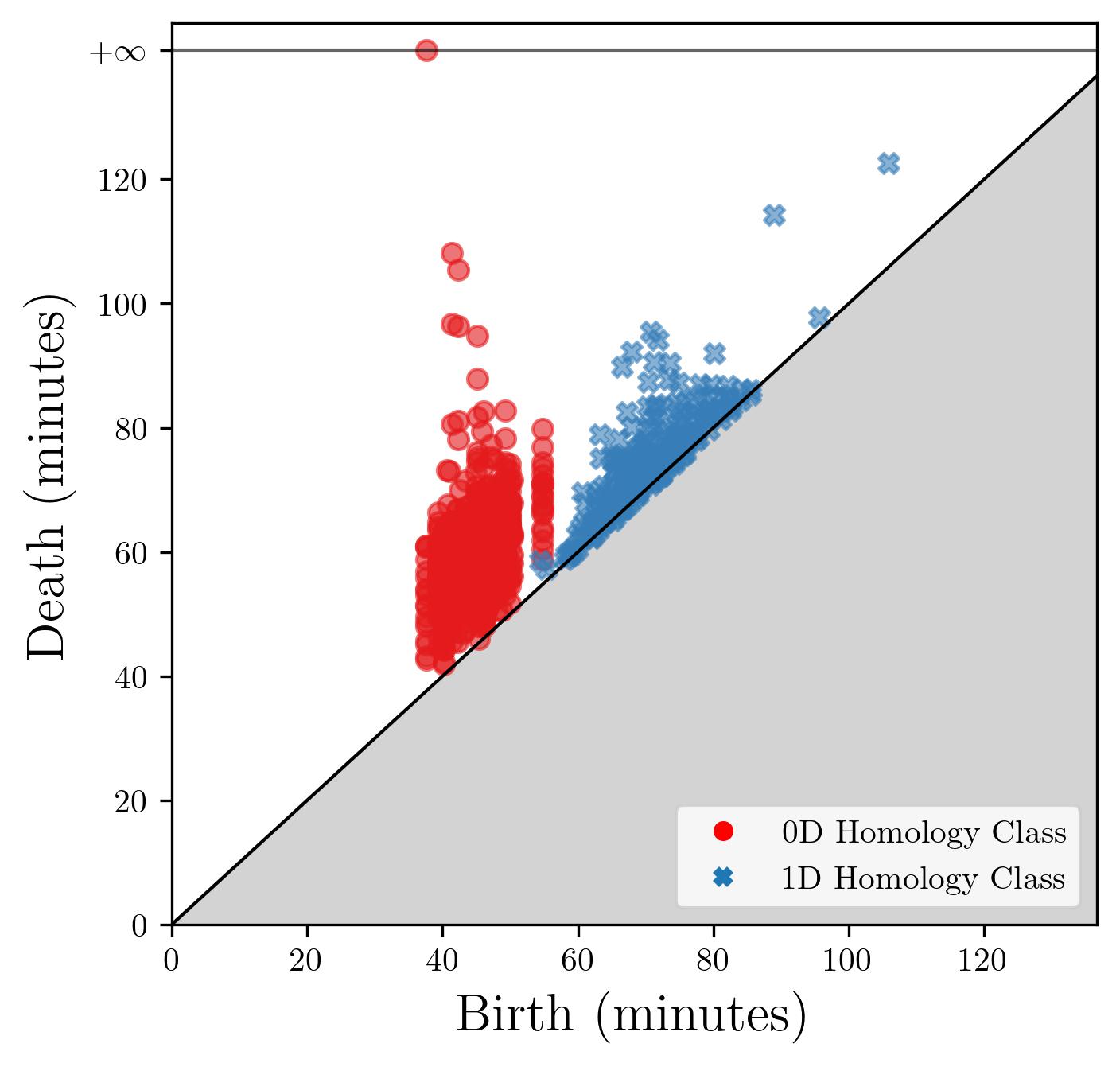}} \\
    \subfloat[New York City]{\includegraphics[width=\halfwidth]{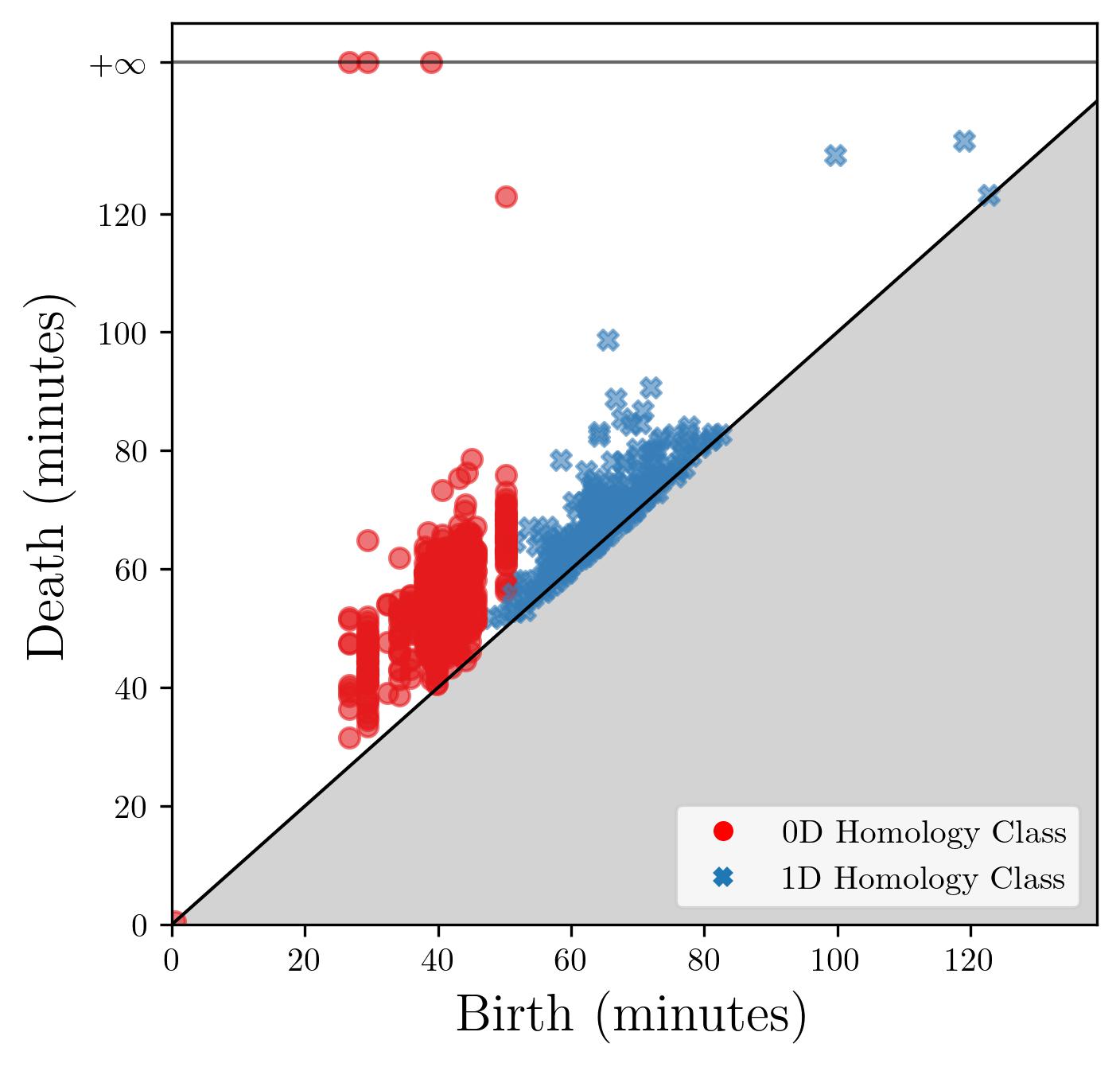}}
    \hspace{.07\textwidth}
    \subfloat[Salt Lake City]{\includegraphics[width=\halfwidth]{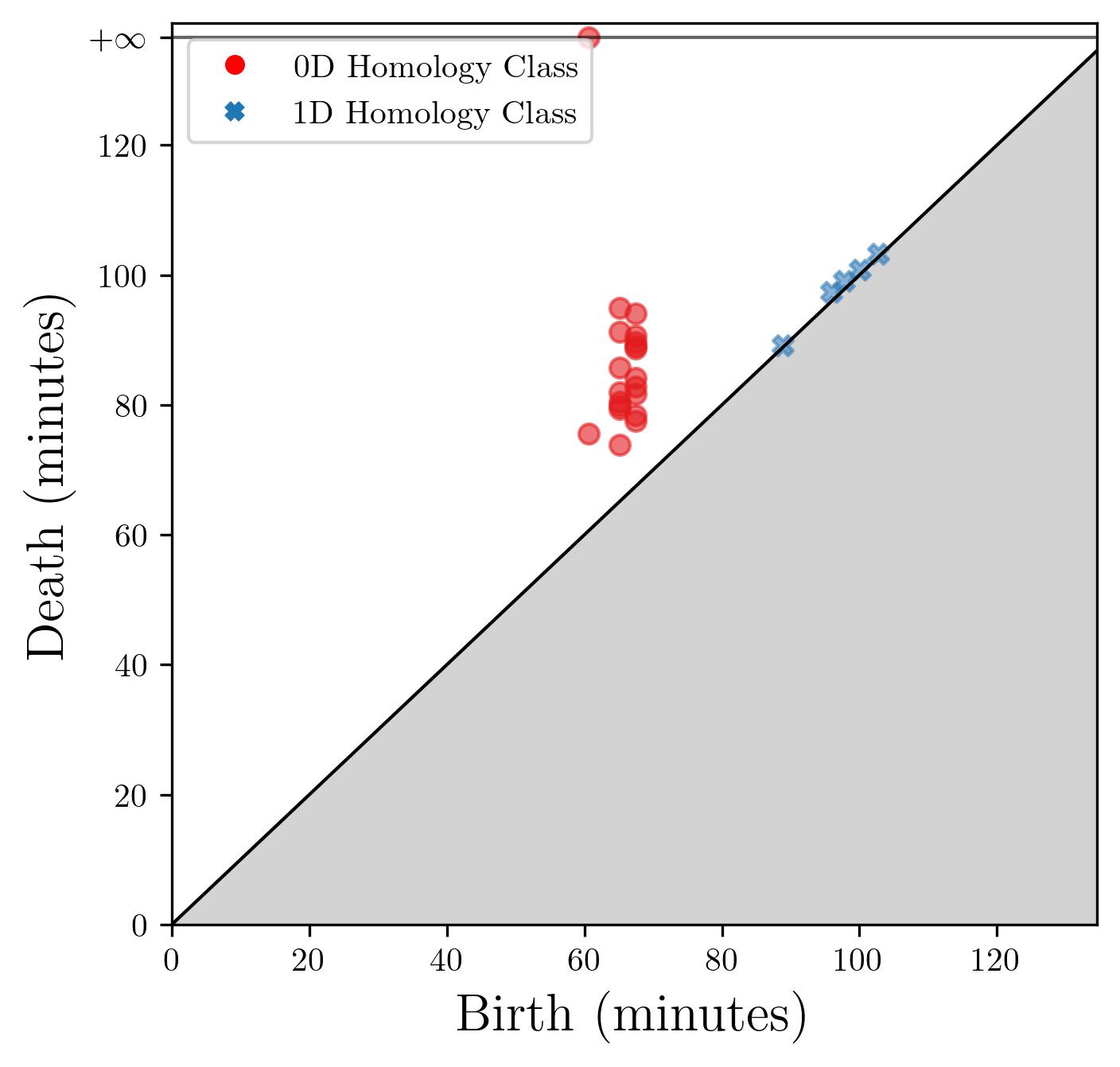}}
    \caption{Our PDs for each city for the PH of the weighted VR complexes that we defined in \cref{sec:methods}.}
    \label{fig:PDs}
\end{figure}

We compute the PH of the weighted VR filtrations of \cref{sec:methods} for Atlanta, Chicago, Jacksonville, Los Angeles County, New York City,
and Salt Lake City. We show their PDs in \cref{fig:PDs}. \major{We examine both 0D and 1D homology classes. The 0D homology classes represent holes between different connected regions of coverage, and the 1D homology classes represent holes in coverage that are bounded by closed paths.} A homology class that dies at filtration-parameter value $t$ represents a hole in coverage that persists until time $t$. One can interpret this to mean that an individual who lives in a hole in coverage that dies at $t$ needs $t$ minutes (including both waiting time at a polling site and travel time back and forth to the site) to cast a vote. 

\major{
In our analysis, we emphasize homology-class death values. We view homology-class birth values as largely irrelevant to our application.
A homology-class birth value indicates the filtration-parameter value at which a coverage hole materializes. 
We use birth values only in the following way. If the death value divided by the birth value (i.e., the ``death/birth ratio") of a homology class is very small (i.e., it is close to $1$), then it is possible that it is merely an artifact of using a VR approximation of a \v{C}ech complex.
We thus focus on homology classes whose death/birth ratios are at least $1.05$.\footnote{
Interested readers can explore thresholds other than 1.05
by using our data, which is available at \url{https://bitbucket.org/jerryluo8/coveragetda/src/main/}. We describe the data in detail in the file ``readme.txt".}
Beyond this, we use only the homology-class death values and death simplices}.

\begin{figure}
    \centering
    \includegraphics[width=\textwidth]{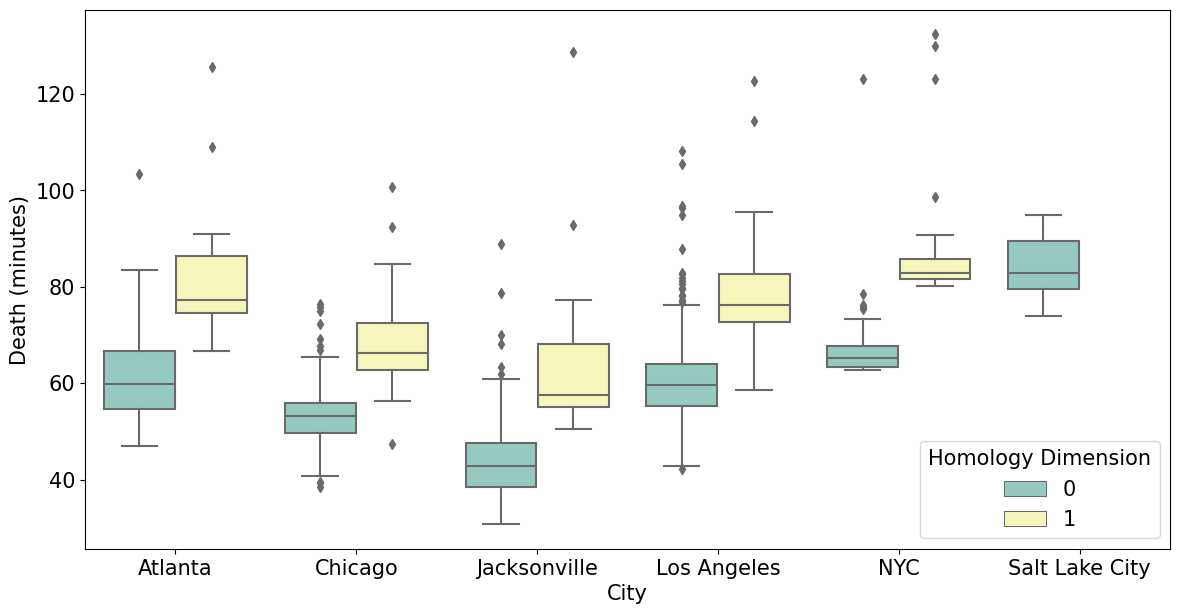}
    \caption{\major{Box plots of the death values of the 0D and 1D homology classes for each city. We only consider homology classes whose death/birth ratio is at least $1.05$.}}
    \label{fig:boxplot}
\end{figure}

Larger homology-class death values suggest that a city may have worse coverage, and a wider distribution of death values suggests that there may be more variation in polling-site accessibility within a city. In \cref{fig:boxplot}, we show a box plot of the distribution of homology-class death values for each city. In \cref{table:stats}, we show the medians and variances of the 0D and 1D homology-class death values for each city.

\begin{table}[]
\centering
\begin{tabular}{|l|l|l|l|}
\hline
\multicolumn{1}{|c|}{City}      & \multicolumn{1}{c|}{\begin{tabular}[c]{@{}c@{}}Homology \\ Dimension\end{tabular}} & \multicolumn{1}{c|}{\begin{tabular}[c]{@{}c@{}}Median\\ (minutes)\end{tabular}} & \multicolumn{1}{c|}{\begin{tabular}[c]{@{}c@{}}Variance\\ (minutes)\end{tabular}} \\ \hline
\hline
\multirow{2}{*}{Atlanta}        & 0                                                                                & \major{59.9}                                                                            & \major{75.4}                                                                            \\ \cline{2-4} 
                                & 1                                                                                  & \major{77.1}                                                                            & \major{150.8}                                                                            \\ \hline \hline
\multirow{2}{*}{Chicago}        & 0                                                                                  & 53.1                                                                           & 30.2                                                                              \\ \cline{2-4} 
                                & 1                                                                                  & \major{66.3}                                                                            & \major{59.7}                                                                              \\ \hline \hline
\multirow{2}{*}{Jacksonville (Florida)}   & 0                                                                                  & 42.8                                                                            & 75.7                                                                              \\ \cline{2-4} 
                                & 1                                                                                  & 57.5                                                                            & \major{394.4}                                                                             \\ \hline \hline
\multirow{2}{*}{Los Angeles County}             & 0                                                                                  & \major{59.6}                                                                            & \major{53.3}                                                                          \\ \cline{2-4} 
                                & 1                                                                                  & \major{76.1}                                                                            & \major{84.6}                                                                             \\ \hline \hline
\multirow{2}{*}{New York City}            & 0                                                                                  & \major{65.1}                                                                            & \major{49.2}                                                                             \\ \cline{2-4} 
                                & 1                                                                                  & \major{82.9}                                                                          & \major{207.1}                                                                               \\ \hline \hline
\multirow{2}{*}{Salt Lake City} & 0                                                                                  & \major{82.8}                                                                           & 37.3                                                                              \\ \cline{2-4} 
                                & 1                                                                                  & \major{N/A}                                                                           & \major{N/A}                                                                              \\ \hline
\end{tabular}
\caption{\major{The medians and variances of the homology-class death values for each city. (As we discussed in the main text, we consider Los Angeles County, rather than only the city of Los Angeles.) We consider homology classes whose death/birth ratio is at least $1.05$. Salt Lake City has no such 1D homology classes.}}
\label{table:stats}
\end{table}

\begin{figure}
    \centering
    \subfloat[\major{0D homology classes}]{\includegraphics[width = .5\textwidth]{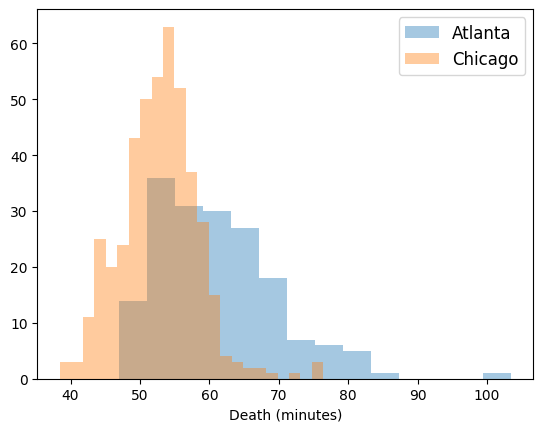}}
    \subfloat[\major{1D homology classes}]{\includegraphics[width = .5\textwidth]{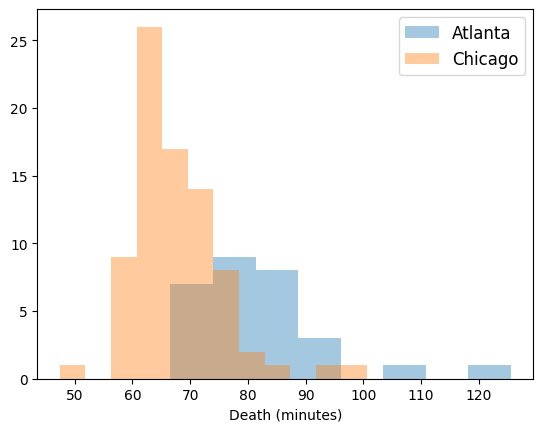}}
    \caption{\major{Histograms of the death values of the 0D and 1D homology classes for Atlanta and Chicago. We only consider homology classes whose death/birth ratio is at least $1.05$.}}
    \label{fig:atl_chc}
\end{figure}

We compare the coverages of the cities by examining the death values in the PDs. For example, in the PDs for Atlanta and Chicago in \cref{fig:PDs}, we see that Atlanta's homology classes tend to die later 
than Chicago's homology classes. We also see this in \cref{fig:boxplot}, in which we show box plots of the death values for each city, and in \cref{fig:atl_chc}, in which we plot the distributions of death values for Atlanta and Chicago. Our PDs and visualizations of summary statistics suggest that Chicago has better polling-site coverage than Atlanta.

\begin{figure}
    \centering

    \subfloat[\major{Atlanta}]{\includegraphics[width = \halfwidth, trim=1in 1.5in 1in 1.5in,clip]{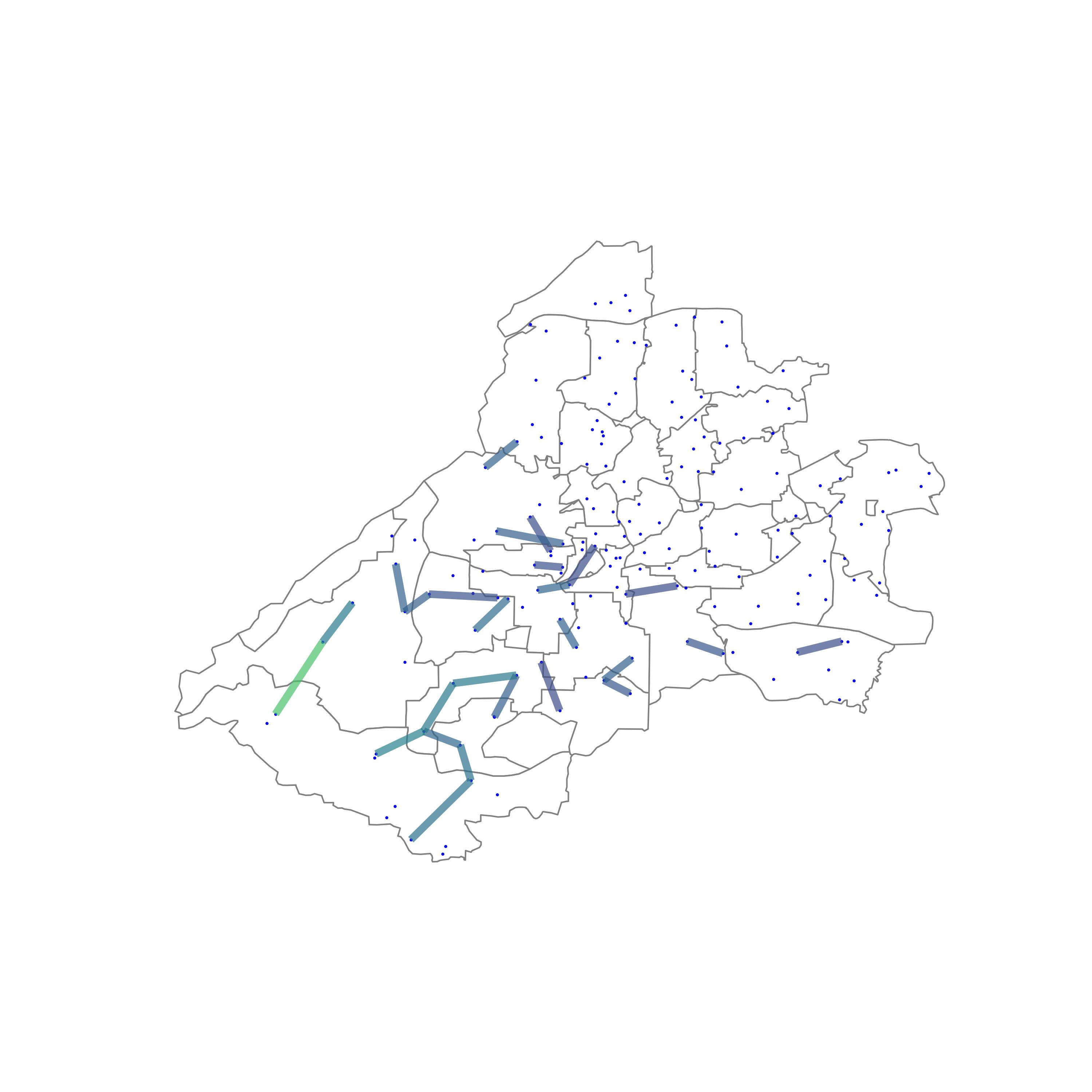}}
    \hspace{.07\textwidth}
    \subfloat[\major{Chicago}]{\includegraphics[width = \halfwidth, trim=1in 1.5in 1in 1.5in,clip]{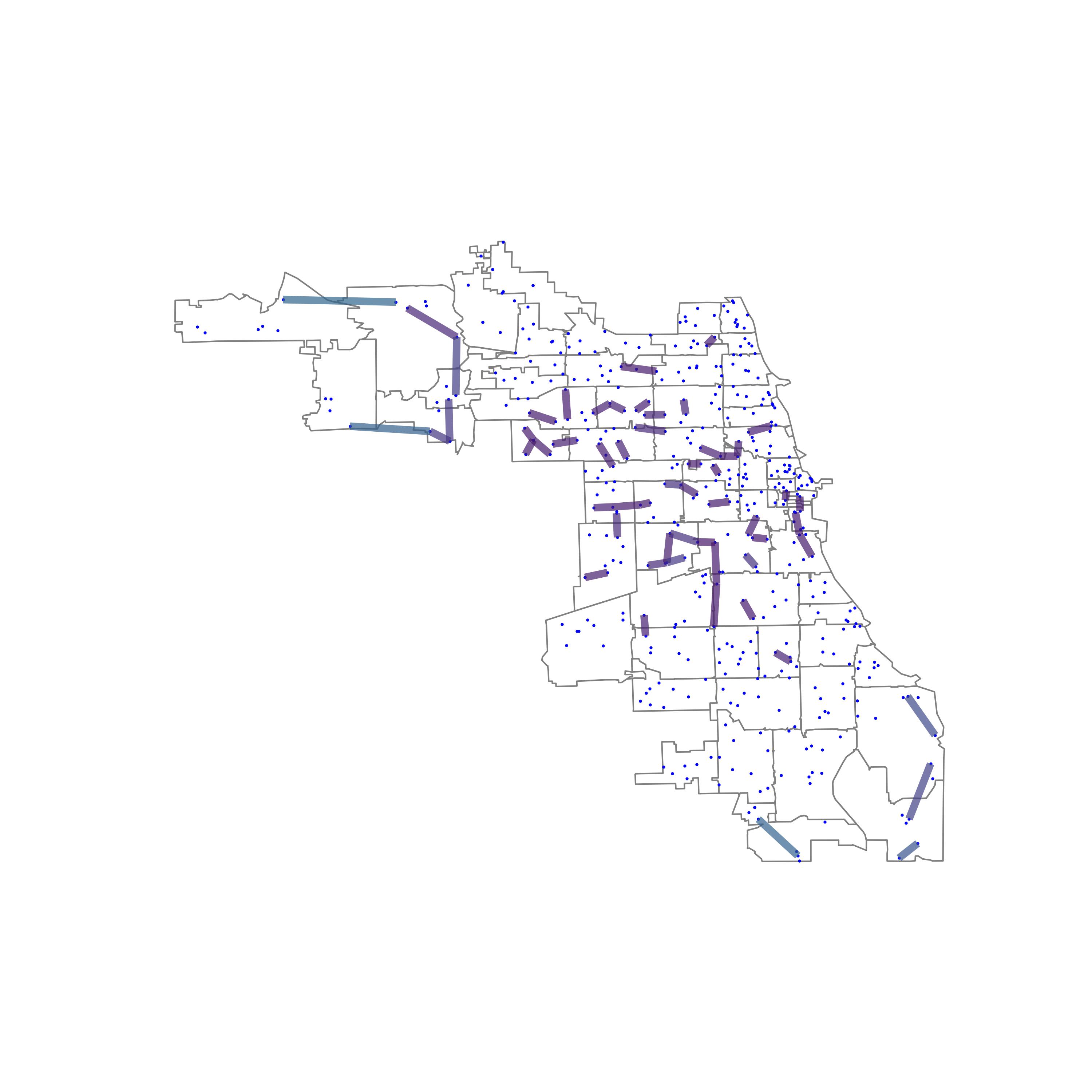}} \\
    \subfloat[\major{Jacksonville (Florida)}]{\includegraphics[width = \halfwidth, trim=1in 1.5in 1in 1.5in,clip]{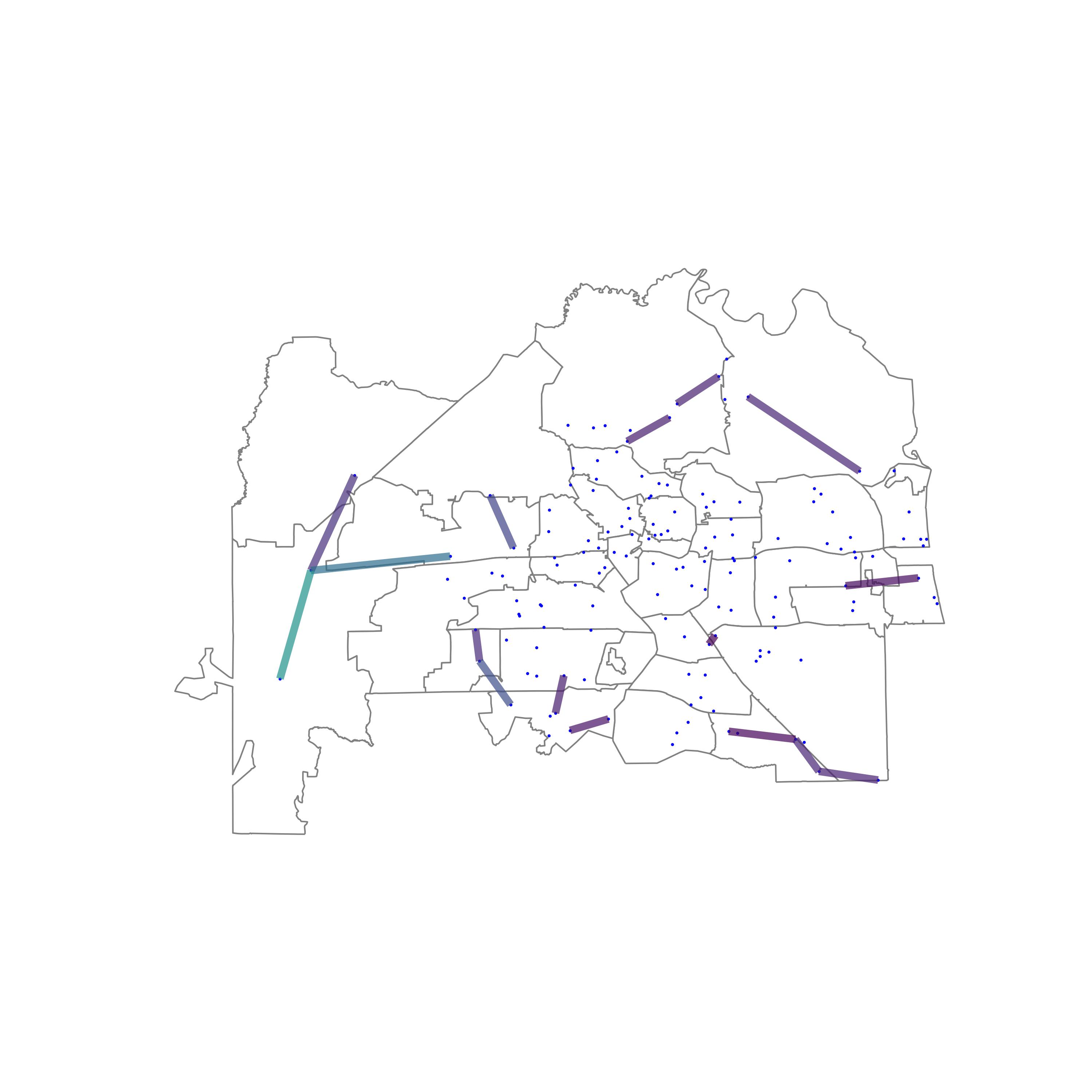}} 
    \hspace{.07\textwidth}
    \subfloat[\major{Los Angeles County}]{\includegraphics[width = \halfwidth, trim=1in 1.5in 1in 1.5in,clip]{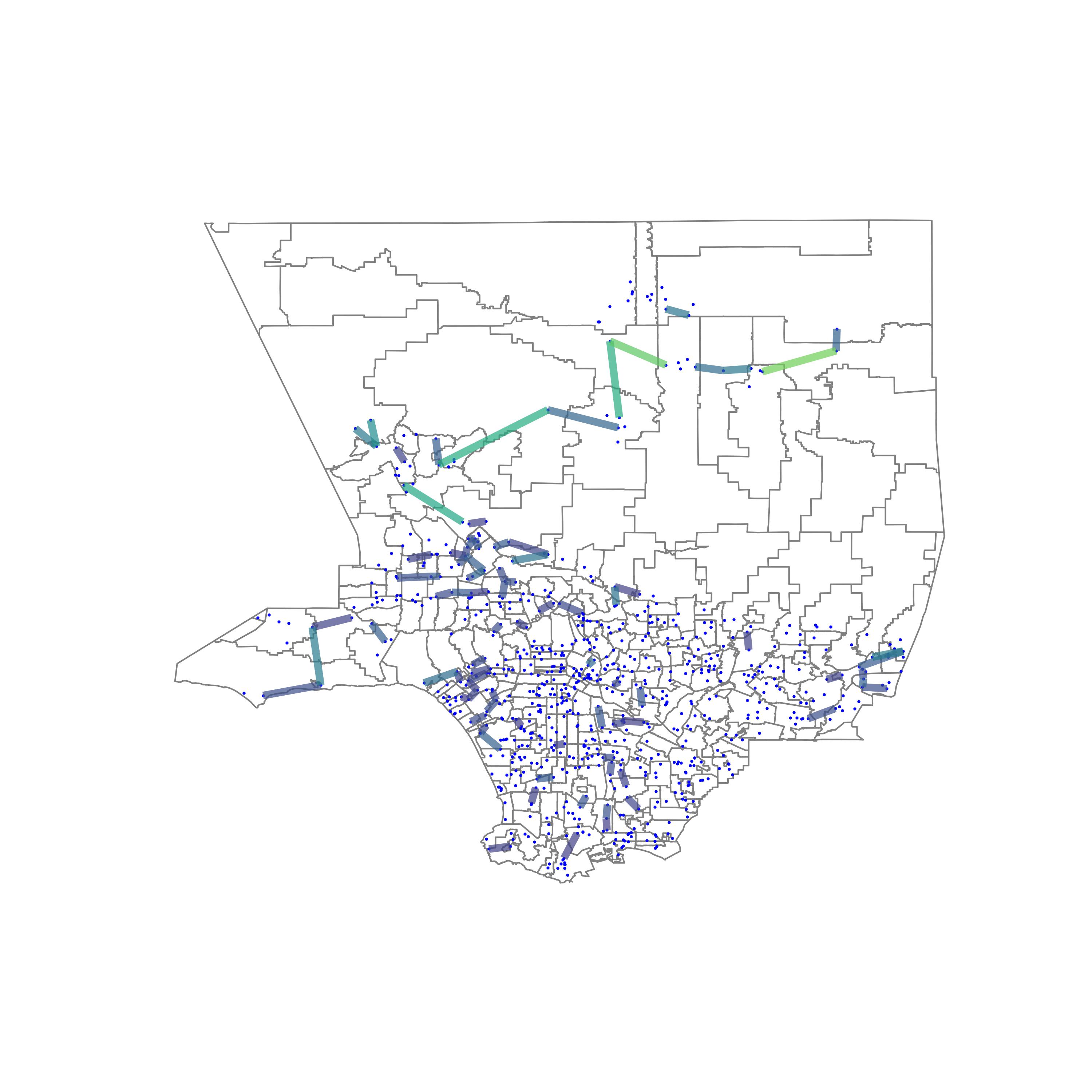}} \\
    \subfloat[\major{New York City}]{\includegraphics[width = \halfwidth, trim=1in 1.5in 1in 1.5in,clip]{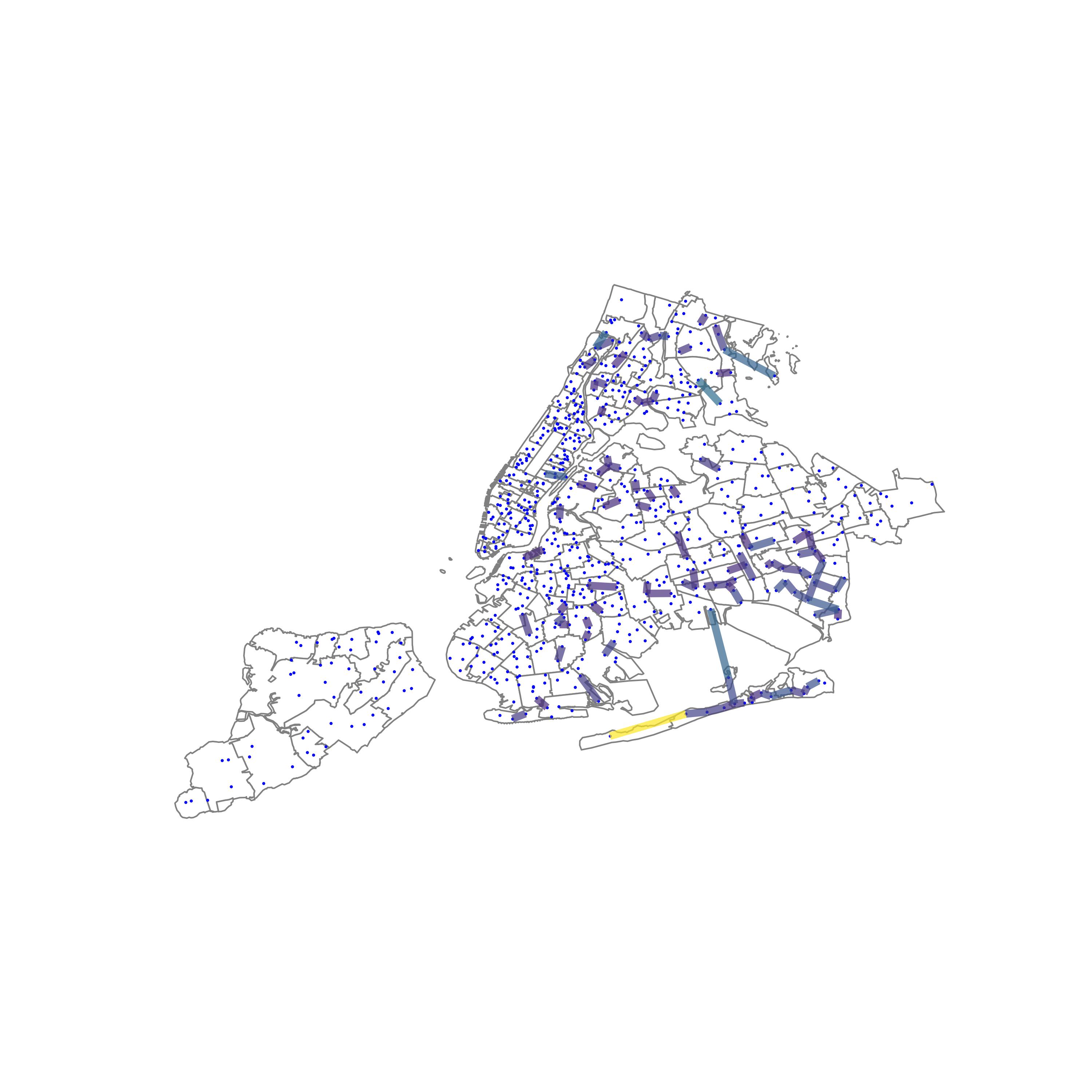}}
    \hspace{.07\textwidth}
    \subfloat[\major{Salt Lake City}]{\includegraphics[width = \halfwidth, trim=1in 1.5in 1in 1.5in,clip]{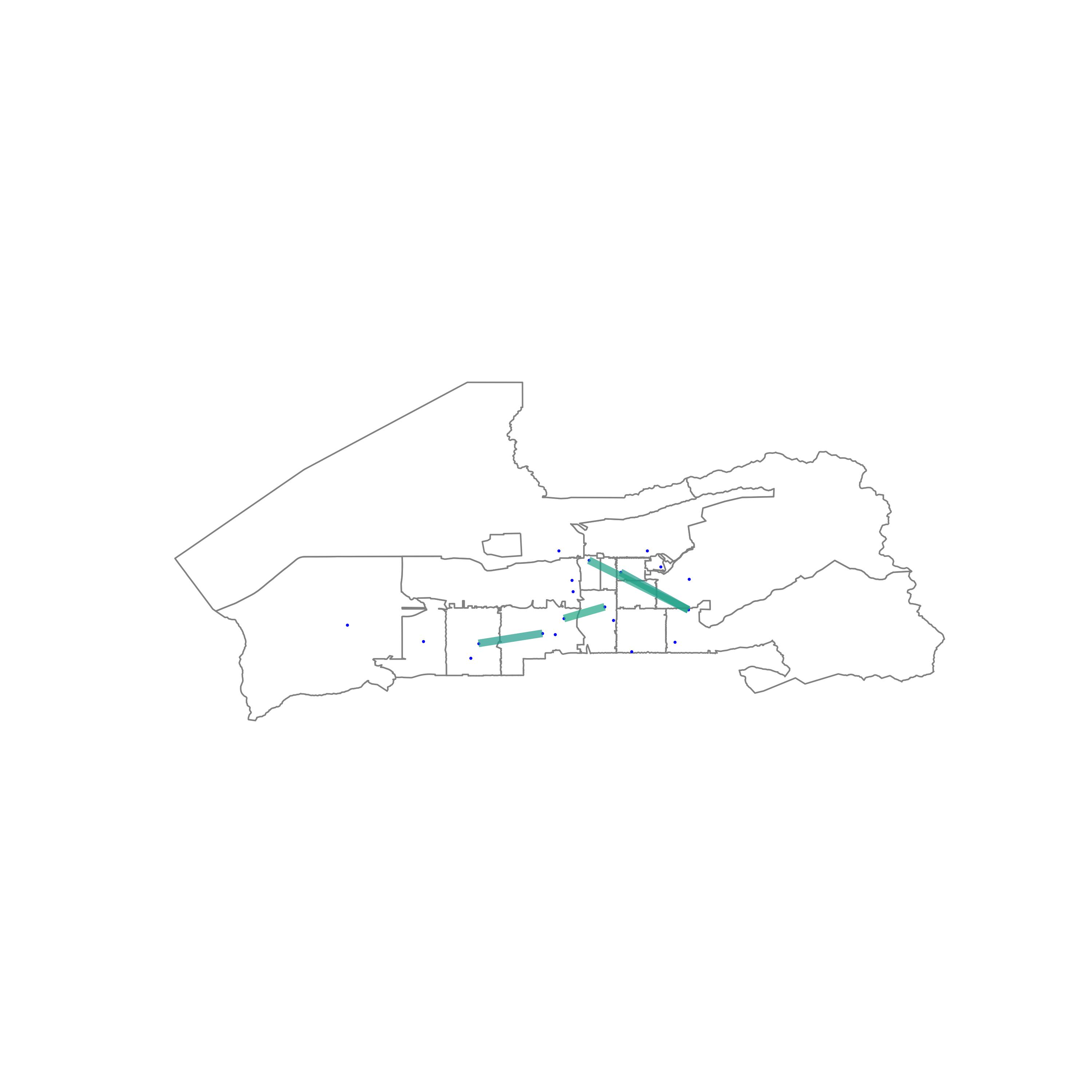}}\\
    \subfloat{\includegraphics[width = \textwidth]{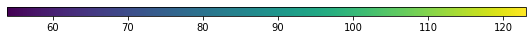}}
    \caption{\major{Death simplices with the largest death values for the 0D homology classes. The colors correspond to the death values (in minutes). We only consider homology classes whose death/birth ratio is at least $1.05$.}
    }
    \label{fig:0D_death_simplices}
\end{figure}

\begin{figure}
    \centering
    \subfloat[\major{Atlanta}]{\includegraphics[width = \halfwidth, trim=1in 1.5in 1in 1.5in,clip]{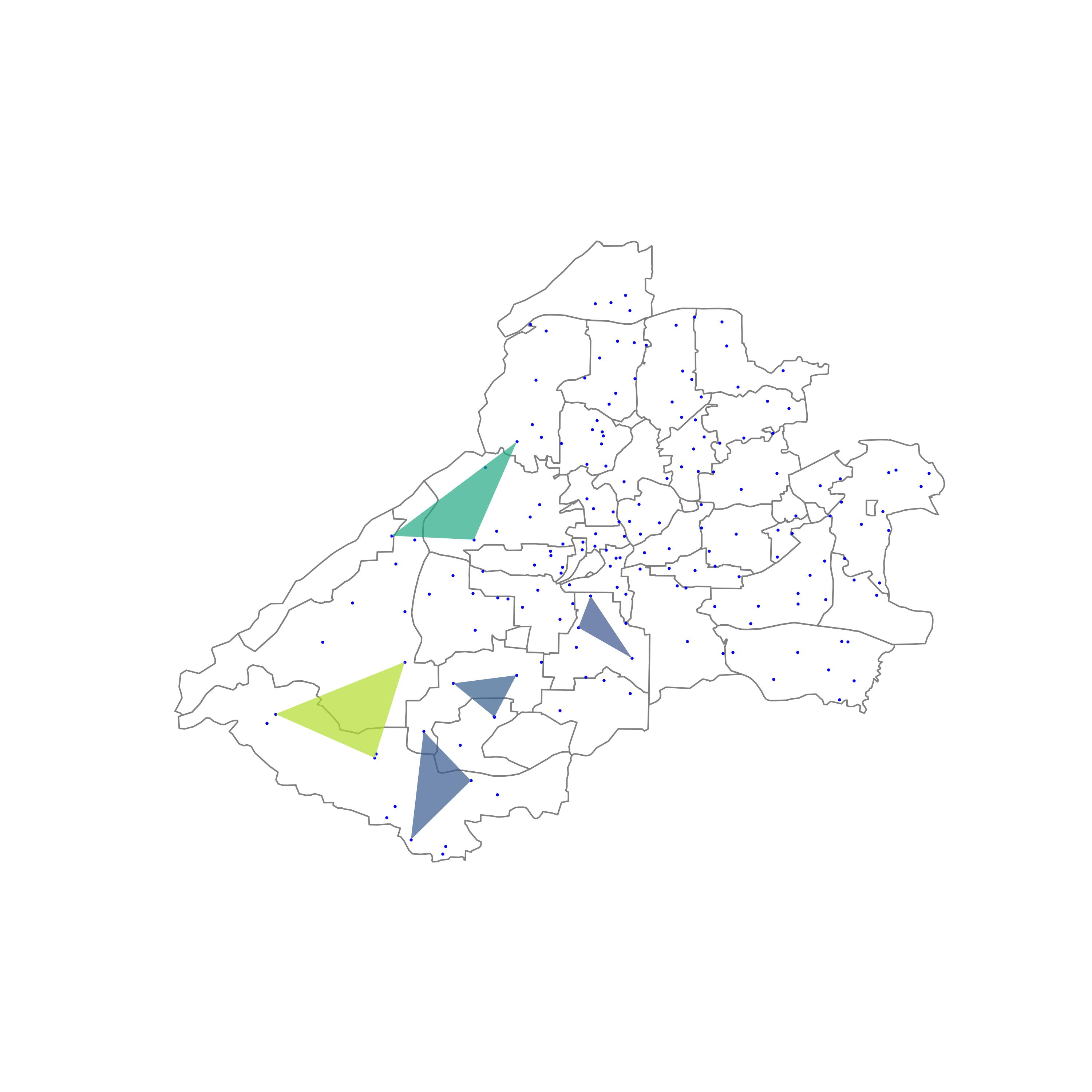}}
    \hspace{.07\textwidth}
    \subfloat[\major{Chicago}]{\includegraphics[width = \halfwidth, trim=1in 1.5in 1in 1.5in,clip]{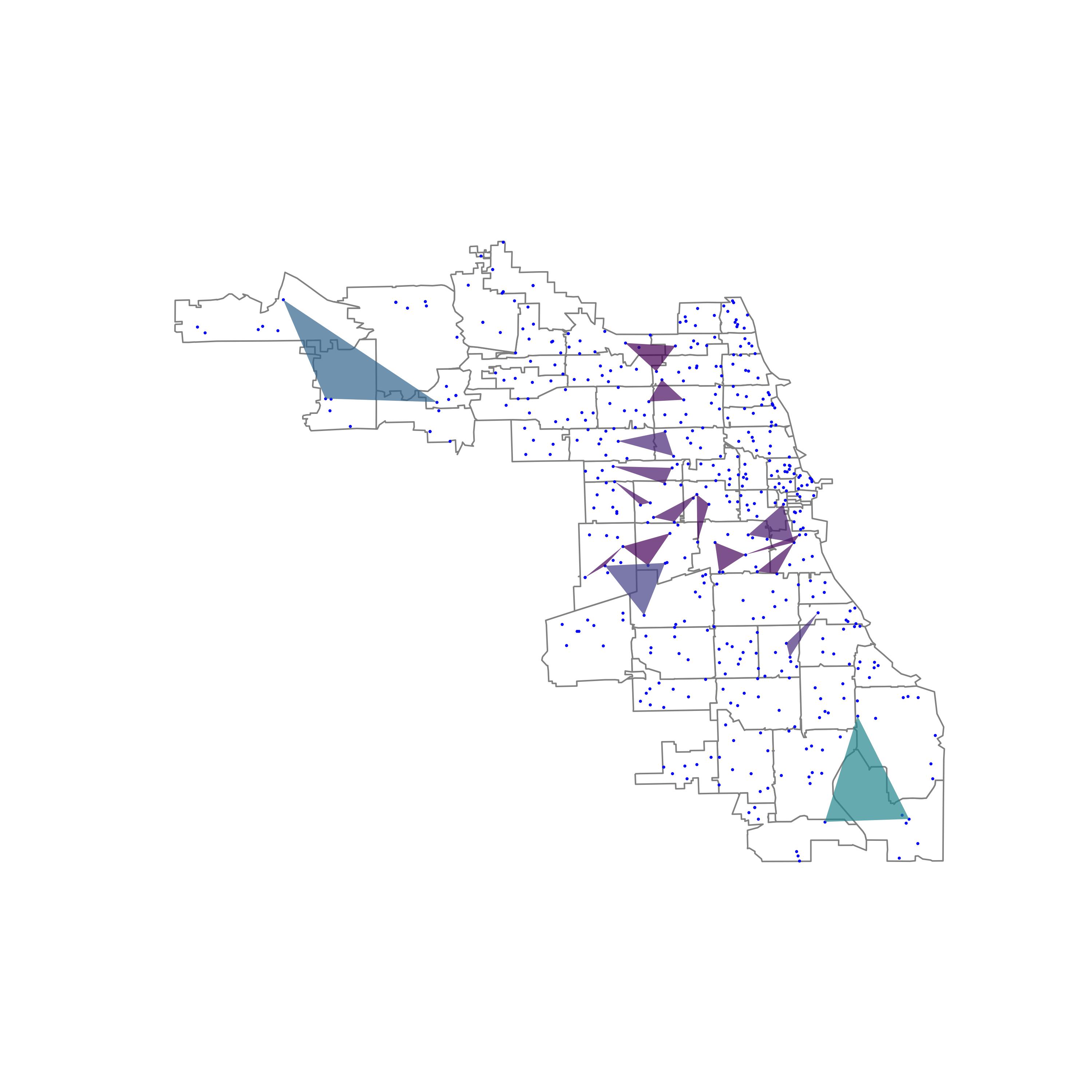}}\\
    \subfloat[\major{Jacksonville (Florida)}]{\includegraphics[width = \halfwidth, trim=1in 1.5in 1in 1.5in,clip]{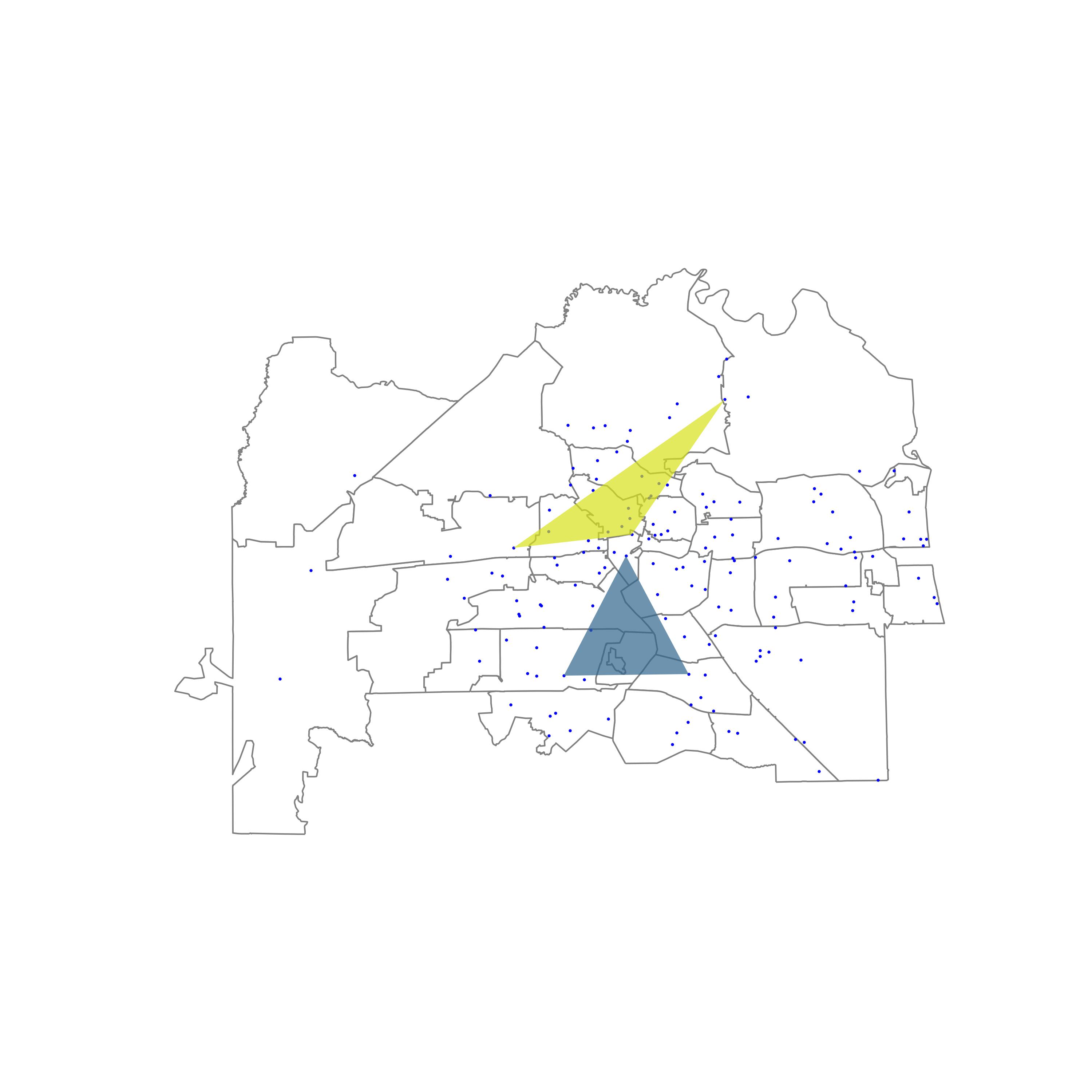}}
    \hspace{.07\textwidth}
    \subfloat[\major{Los Angeles County}]{\includegraphics[width = \halfwidth, trim=1in 1.5in 1in 1.5in,clip]{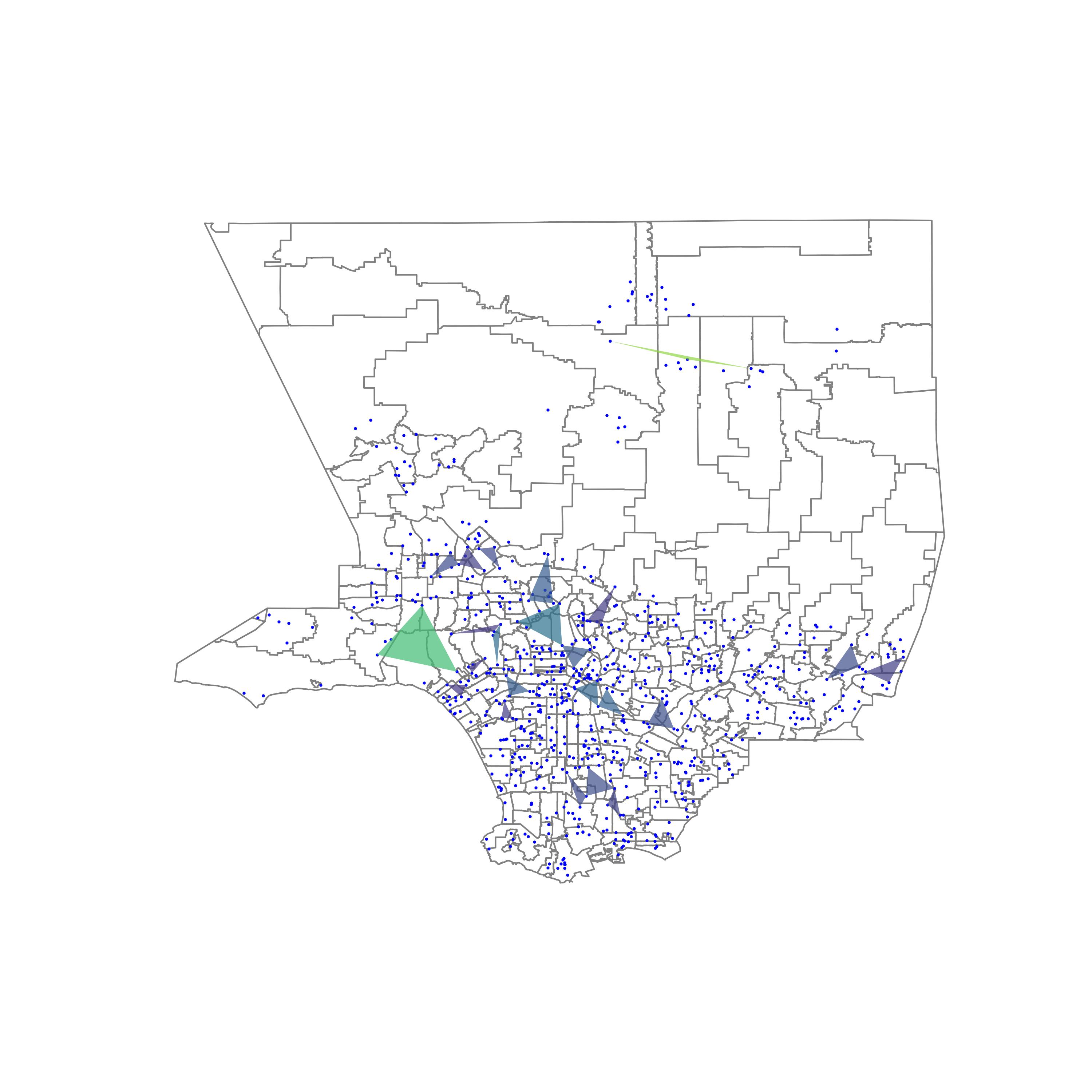}} \\
    \subfloat[\major{New York City}]{\includegraphics[width = \halfwidth, trim=1in 1.5in 1in 1.5in,clip]{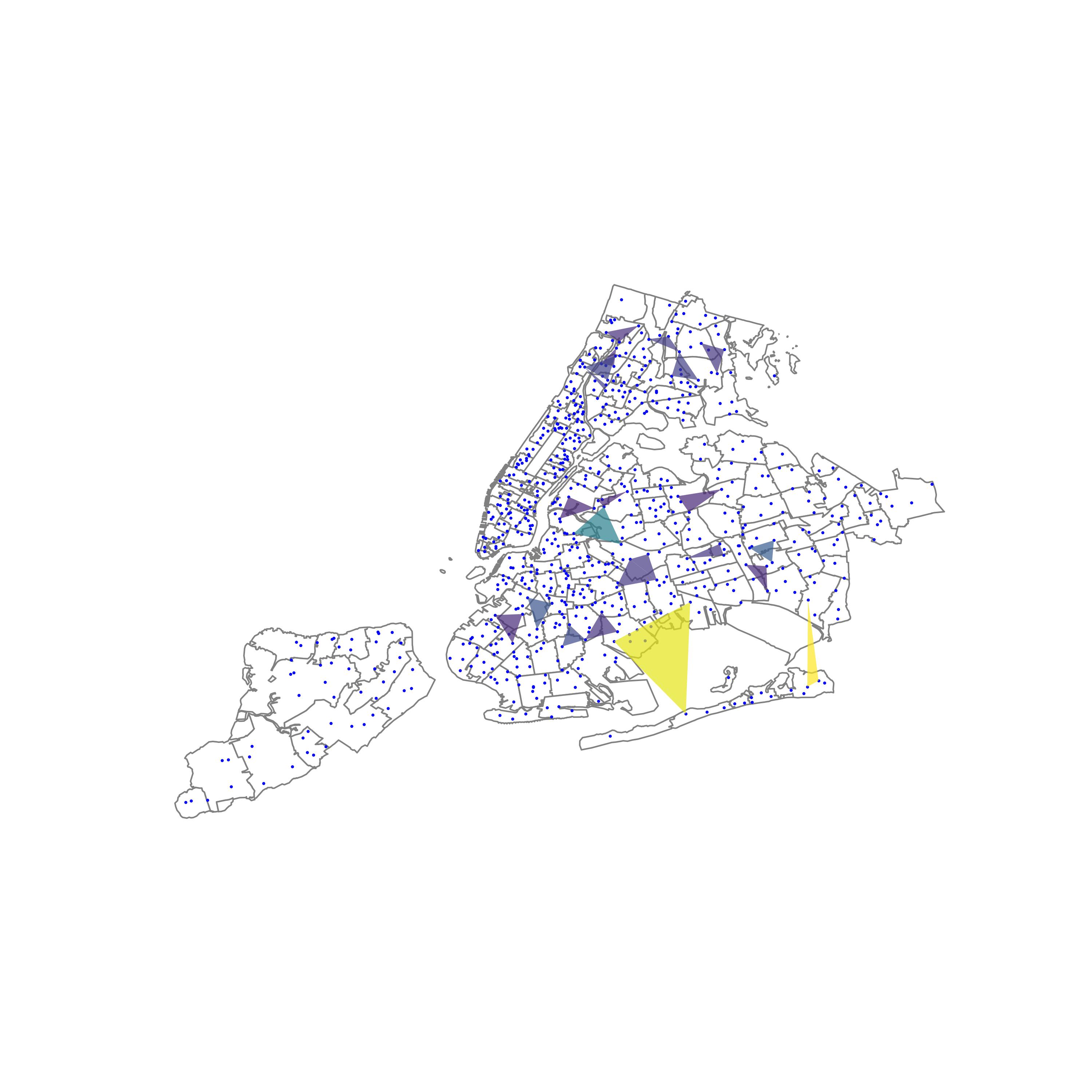}}
    \hspace{.07\textwidth}
    \subfloat[\major{Salt Lake City}]{\includegraphics[width = \halfwidth, trim=1in 1.5in 1in 1.5in,clip]{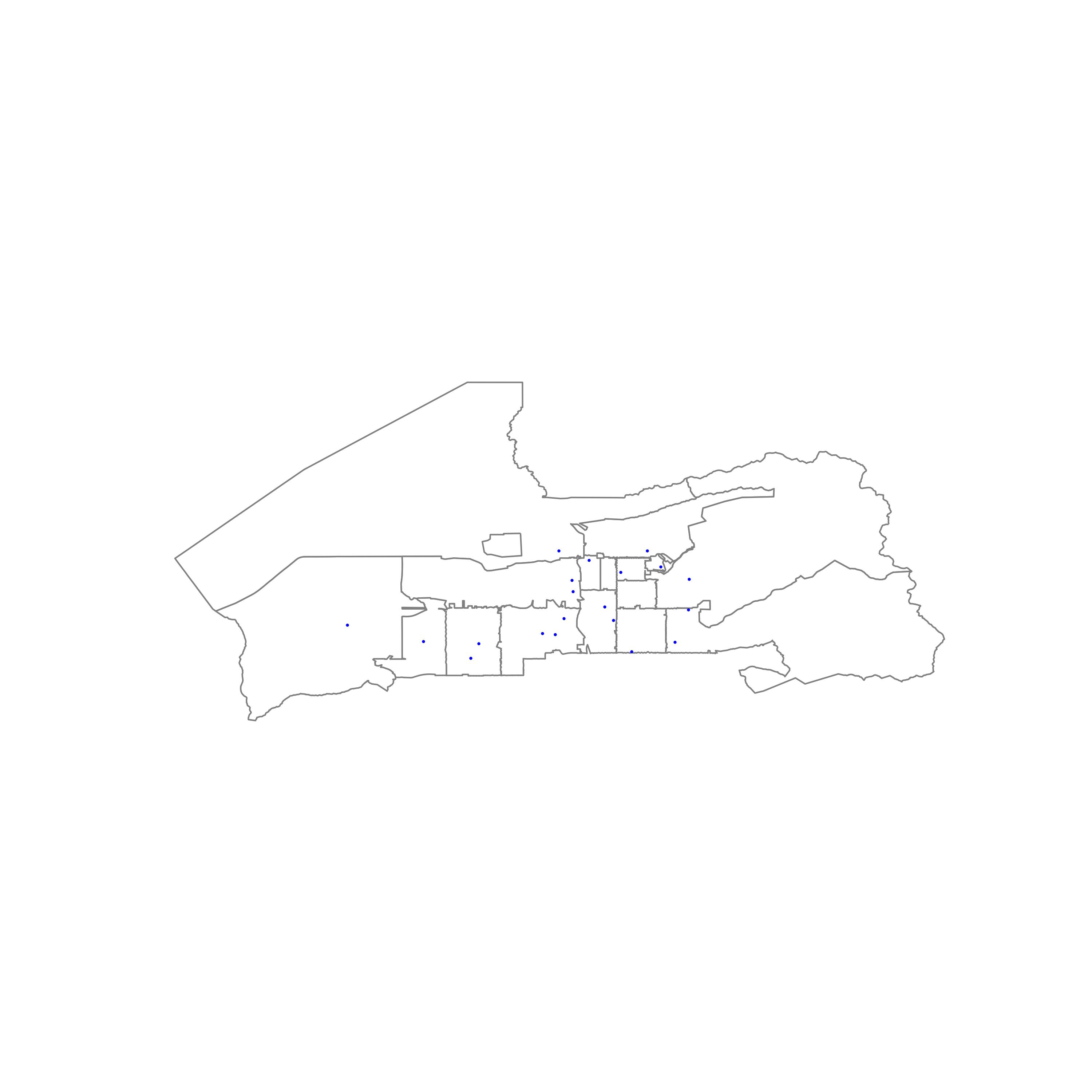}}\\
    \subfloat{\includegraphics[width=\textwidth]{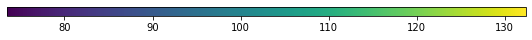}}
    
    \caption{\major{Death simplices with the largest death values for the 1D homology classes. The colors correspond to the death values (in minutes). We only consider homology classes whose death/birth ratio is at least $1.05$.}}\label{fig:1D_death_simplices}
\end{figure}

We use the death simplices to locate and visualize holes in polling-site coverage. We interpret the death simplex of a homology class as the ``epicenter'' of an associated coverage hole because the death simplex represents the last part of the hole to be covered. The death simplex of a 0D homology class is an edge between two polling sites; there is a hole in coverage between those two sites. Similarly, the death simplex of a 1D homology class is a triangle that is the convex hull of three polling sites; there is a hole in coverage between those three sites.
In Figures \ref{fig:0D_death_simplices} and \ref{fig:1D_death_simplices}, we show the death simplices with the largest death values\footnote{More precisely, for each city and each homology dimension ($0$ and $1$), we show the death simplices whose death values have a z-score of at least $1$. We calculate the z-score as follows. Let $d$ be the death value of a {$\homdim$}-dimensional homology class (where $\homdim=0$ or $\homdim=1$) for city $C$. The z-score of $d$ is $z = (d - \mu_{C, \homdim})/\sigma_{C, \homdim}$, where $\mu_{C, \homdim}$ and $\sigma_{C, \homdim}$ are the mean and standard deviation of the distribution of death values of the $\homdim$-dimensional homology classes for city $C$.} for the 0D and 1D homology classes,\footnote{\major{In \Cref{fig:1D_death_simplices}, in which we show the death simplices of the 1D homology classes, some of the polling sites appear to be covered by death simplices whose vertices are other polling sites.
} 
\major{At least two factors may contribute to this. One factor is that our measure of distance is not a Euclidean metric, even though we plot the death simplices in \Cref{fig:1D_death_simplices} as Euclidean triangles. The Euclidean triangles can sometimes cover polling sites that are not among its vertices, but geodesic triangles may not cover those polling sites. 
Another possibility is that a polling site $x$ has such a long waiting time that it does not show up in the filtration until after the homology class whose death simplex includes $x$ has already died.}
} 
respectively. For example, consider panels (a) and (b) of \cref{fig:0D_death_simplices} and \cref{fig:1D_death_simplices}, in which we show the 0D and 1D homology-class death simplices for Atlanta and Chicago. The areas of lowest coverage (i.e., the areas that have the death simplices with the largest death values) in Atlanta tend to be in the southwest, whereas the areas of lowest coverage in Chicago tend to be in the northwest and southeast. There is one homology class in Atlanta that has a significantly larger death filtration value than the other classes in Atlanta and any of the classes in Chicago. This homology class represents a 1D hole in coverage in southwest Atlanta (see \cref{fig:1D_death_simplices}a).

\section{Conclusions and Discussion}\label{sec:discussion}

\subsection{Summary}

We showed that persistent homology (PH), which is a type of topological data analysis (TDA), is a helpful approach to study accessibility and equitability. It allows one to examine holes in resource coverage with respect to an appropriate choice of ``distance'', which one constructs to incorporate important features of a problem of interest. The distance can be based on geography, time, or something else. In the present 
paper, we used PH to study and quantify holes in polling-site coverage in six United States cities (technically, in five cities and in Los Angeles County). For each city, we constructed a filtration in which a homology class that dies at time $t$ represents a geographical region in which it takes $t$ minutes to cast a vote (including both travel time and waiting time). We interpreted the death simplex of a homology class as the location of the corresponding hole in resource coverage. The information in the PH allowed us both to compare the accessibility of voting across our chosen cities and to determine the locations of the coverage holes within each city.

A key benefit of our use of PH is that it enabled us to identify holes in polling-site coverage at all time scales. It also 
allowed us to use a distance that we designed for the problem at hand, rather than merely using geographical distance, which does not capture important factors in resource accessibility~\cite{brabyn_beere}. We based our distance function on estimates of travel time, which is more reasonable and accurate than geographical distance for capturing resource accessibility~\cite{Pearce2006-vf}.


\subsection{Limitations}\label{sec:limitations}
To conduct our study, we needed to estimate a variety of quantities (see \cref{sec:methods}), including travel times, waiting times, and demographic information. We also made several simplifications because of computational and monetary constraints. We now discuss some issues that are important to address before attempting to incorporate our approach into policy-making. 

One limitation of our study is our estimation of travel times. As we discussed in \cref{sec:traveltime}, we computed travel times using the Google Maps API. Because of monetary constraints, we only computed a subset of the relevant travel times and used a graph-based estimate to determine the others. Additionally, we computed each travel time between polling sites only once. Computing more precise estimates of travel times is important to better capture the accessibility of polling sites. One way to do this is to compute the same travel time multiple times across different days and times of day 
and take an average. Such additional computations can also help yield estimates of best-case and worst-case scenarios.

Another limitation of our study is the granularity of our data. As we discussed in \cref{sec:waittime}, our waiting-time data is at the scale of congressional districts. Because there is heterogeneity in the waiting times at different polling sites in the same congressional district, it is important to obtain finer-grained data for the waiting times at polling sites. Having finer-grained waiting times (e.g., if possible, procuring an estimated waiting time for each polling site) would improve our ability to capture voting accessibility.

We also made several topological approximations. We worked with a weighted VR filtration, which approximates a weighted \v{C}ech filtration, which in turn approximates the nested set $\{\bigcup B(x_i, r_{x_i}(t))\}_{t \in \mathbb{R}}$ of spaces, where $\{x_i\}$ is a set of polling-site locations and $r_{x_i}(t)$ is the radius function that we defined in \cref{sec:methods}. The nested set of spaces is directly relevant to our application, as the holes in $\bigcup B(x_i, r_{x_i}(t))$ are the true holes in polling-site coverage. We made our approximations, which are standard in TDA and are well-justified (see our discussion in \cref{sec:background}) \cite{roadmap}, to reduce computational cost. However, the convexity condition of the Nerve Theorem, which justifies the approximation of $\bigcup B(x_i, r_{x_i}(t))$ by a weighted \v{C}ech complex, is not guaranteed to be satisfied for all times $t$. The Nerve Theorem implies that the weighted \v{C}ech complex is homotopy-equivalent to $\bigcup B(x_i, r_{x_i}(t))$ whenever the balls $B(x_i, r_{x_i}(t))$ are convex. This condition always holds in Euclidean space, but it is not guaranteed to hold in the space that we defined in \cref{sec:methods}.\footnote{\major{Although our space is not Euclidean, it is still reasonable to assume that it is approximately \emph{locally} Euclidean. That is, for each polling site $x$, there is a constant $a > 0$ such that if $y$ is sufficiently close to $x$, then $d(x, y) \approx a \cdot d_E(x, y)$, where $d(x, y)$ is defined by \cref{eq:distance} and $d_E(x, y)$ is the Euclidean distance. This approximation holds because car-ownership rates and traffic conditions do not vary much within a sufficiently small neighborhood.} 
\major{We verified empirically that our distance function is approximately locally Euclidean by showing that, for each polling site $x$, there is a strong linear correlation between the pairwise distances $d(x, y)$ and the pairwise Euclidean distances $d_E(x, y)$ when $y$ is sufficiently close to $x$. Because our distance function is approximately locally Euclidean, sufficiently small balls (with respect to our distance function)
behave like Euclidean balls, so the Nerve Theorem is applicable for sufficiently small filtration values.}} 
Homotopy-equivalence is important because homotopy-equivalent
spaces have the same homology and thus have the same set of holes.

Finally, our approach only detects holes in the convex hull of a set of resource sites. Although this may be inconsequential if resource sites are sufficiently spread out geographically, it can be problematic if the resource sites are overly concentrated near a few locations. One way to address this issue is to incorporate 
city boundaries into the construction of the filtrations. This would help capture holes in coverage in regions that lie outside the convex hull of the resource sites, and it would also help identify the filtration-parameter value $t$ at which an entire city is covered by the balls $B(x_i, r_{x_i}(t))$.


\subsection{Future work}\label{sec:future}

As we discussed in \cref{sec:limitations}, we made several topological approximations of our mathematical object of interest, which is the nested set $\{\bigcup B(x_i, r_{x_i}(t))\}_{t \in \mathbb{R}}$ of spaces. Instead of using a weighted VR filtration, one can construct a more direct approximation of $\{\bigcup B(x_i, r_{x_i}(t))\}_{t \in \mathbb{R}}$. One can first discretize a city by imposing a grid onto it. For each point on such a grid, one can then construct the filtered cubical complex that is induced by the distance to the nearest polling site. However, this is much more computationally expensive than our approach, and it would also entail many more travel-time queries (which cost money) than in the present paper.\footnote{Our distance function \cref{eq:distance} is symmetric, but recall that it is not a metric because it does not satisfy the triangle inequality. Therefore, we cannot use techniques such as distance transforms and level-set propagation to reduce the computational complexity of calculating the filtration $\{\bigcup_i B(x_i, r_{x_i}(t))\}_{t \in \mathbb{R}}\}$.}

It is also important to incorporate city boundaries into the construction of filtrations. One way to do this is as follows. Let $x_1,\ldots, x_n$ denote the resource sites, and let $y_1,\ldots, y_m$ denote the points that one obtains by discretizing a city boundary. One can extend our distance function \cref{eq:distance} by defining\footnote{The factor of $2$ comes from the fact that $x_i$ is a resource site but $y_j$ is not a resource site.}
\begin{equation}\label{eq:dist_bdry_to_poll}
	d(x_i,y_j) := \frac{2}{ P}[P_{Z(x_i)}\tilde{d}(x_i, y_j) + P_{Z(y_j)} \tilde{d}(y_j, x_i)]\,, 
\end{equation}
where $P$, $Z$, and $\tilde{d}$ are as in the distance function \cref{eq:distance} and
\begin{equation}\label{eq:dist_bdry_to_bdry}
	d(y_i,y_j) = \begin{cases} 
0\,, & y_i \textrm{ and } y_j \textrm{ are adjacent points of the discretized city boundary} \\ 
\infty\,, & \textrm{otherwise\,.}
\end{cases}    
\end{equation}
At each filtration-parameter value, the simplicial complex that one constructs using the distance function \cref{eq:distance} with the extensions \cref{eq:dist_bdry_to_poll} and \cref{eq:dist_bdry_to_bdry} includes both the points that one obtains by discretizing the boundary and the edges that connect adjacent boundary points. The largest death value is then the filtration-parameter value $t$ that corresponds to the time at which
an entire city is covered by the balls $\{B(x_i, r_{x_i}(t))\}$ (i.e., when there are no longer any holes in coverage).

\major{In our paper, we used death simplices to locate holes in coverage, but other approaches are also possible. 
For example, by calculating minimal generators \cite{minimal_cycle}, one can identify representative cycles that encircle holes.
The topological pipeline ``hyperTDA" was introduced recently \cite{hypertda} to analyze the structure of minimal generators by constructing a hypergraph, calculating hypergraph centrality measures, and employing community detection. This approach may provide insights into the spatial structure of minimal generators. Another potentially viable approach is to use decorated merge trees (DMTs) \cite{dmt} to locate 1D holes in coverage. DMTs allow one to match 1D holes with an associated cluster of points.}

Although we explored a specific case study (namely, the accessibility of polling sites), it is also relevant to conduct similar investigations for other resources, such as public parks, hospitals, vaccine distribution centers, grocery stores, Planned Parenthood clinics, and Department of Motor Vehicle (DMV) locations. One can use similar data to construct a filtration, although it may be necessary to modify the choices of distance and weighting. One can also use ideas from mobility theory \cite{BARBOSA20181} to help construct suitable distances and weightings.
For example, all DMV offices offer largely the same services, so it seems reasonable to assume that people will go to their nearest office. Therefore, in a study of DMV accessibility, it seems appropriate to use travel time 
as a distance function, just as we did in our analysis of polling sites. However, in other applications, it is not reasonable to use travel time alone as a distance function. For example, different grocery stores may offer different products at different prices, so travel time alone may not be appropriate as a choice of distance function. 
Additionally, although waiting time is a significant factor for investigating the coverage of polling sites, there are many applications for which it does not make sense to incorporate waiting time. For example, the time that is spent in a public park or recreation center is typically not a barrier to access. In applications for which waiting time is not an accessibility factor, it seems more appropriate to use a standard VR filtration than a weighted VR filtration. With salient modifications (such as the ones that we described in this subsection and in \cref{sec:limitations}), we can apply our approach to many other types of resource sites.


\section*{Acknowledgements}

We thank Chris Anderson and Renata Turkes for helpful comments and discussions. We also thank two anonymous referees for helpful comments on an earlier version of this paper.



\end{document}